\documentstyle[12pt,aaspp4,flushrt]{article}
\tighten
\newcommand{\etal}{{\rm et al.~}}
\newcommand{\Mpc}{$h^{-1}$~{\rm Mpc}}
\newcommand{\hmpc}{$h$~{\rm Mpc$^{-1}$}}

\begin{document}

\title{Steps toward the power spectrum of matter. 
I. The mean spectrum of galaxies}

\author{J. Einasto \altaffilmark{1}, 
{M. Einasto}\altaffilmark{1},
{E. Tago}\altaffilmark{1},
{A. A. Starobinsky}\altaffilmark{2},
{F. Atrio-Barandela}\altaffilmark{3},
{V. M\"uller}\altaffilmark{4}, 
{A. Knebe}\altaffilmark{4},
{P. Frisch}\altaffilmark{5},
{R. Cen}\altaffilmark{6},
{H. Andernach}\altaffilmark{7} and
{D. Tucker}\altaffilmark{8}}

\altaffiltext{1}{Tartu Observatory, EE-2444 T\~oravere, Estonia}
\altaffiltext{2}{Landau Institute for Theoretical Physics, Moscow 117334, 
Russia} 
\altaffiltext{3}{F{\'\i}sica Te\'orica, Universidad de Salamanca, 37008 
Spain } 
\altaffiltext{4}{Astrophysical Institute Potsdam, An der Sternwarte 16,
       D-14482 Potsdam, Germany}
\altaffiltext{5}{G\"ottingen University Observatory, Geismarlandstr. 11,
       D-37083 G\"ottingen, Germany}
\altaffiltext{6}{Department of Astrophysical Sciences, Princeton University,
Princeton, NJ 08544, USA }
\altaffiltext{7}{Depto. de Astronom\'\i a, IFUG,
 Apdo. Postal 144, C.P. 36000 Guanajuato, Gto., Mexico}
\altaffiltext{8}{Fermilab, MS 127, Box 500, Batavia, IL 60510, USA}

\begin{abstract}

  We calculate the mean power spectrum of all galaxies using published
  power spectra of galaxies and clusters of galaxies.  On small scales
  we use the power spectrum derived from the 2-dimensional
  distribution of APM galaxies, since this sample is not influenced by
  redshift distortions and is the largest and deepest sample of
  galaxies available.  On large scales we use power spectra derived
  from 3-dimensional data for various galaxy and cluster samples which
  are reduced to real space and in amplitude to the power spectrum
  of APM galaxies.  We find that available data indicate the presence
  of two different populations in the nearby Universe. Clusters of
  galaxies sample a relatively large region in the Universe where
  rich, medium and poor superclusters are well represented.  Their
  mean power spectrum has a spike at wavenumber $k=0.05 \pm
  0.01$~\hmpc, followed by an approximate power-law spectrum of index
  $n\approx -1.9$ towards small scales.  Some galaxy surveys (APM 3-D,
  IRAS QDOT, and SSRS+CfA2~130 Mpc) have similar spectra.  The power
  spectrum found from LCRS and IRAS 1.2 Jy surveys is flatter around
  the maximum, which may represent regions of the Universe with
  medium-rich and poor superclusters.  Differences in power spectra
  for these populations may partly be due to the survey geometries of
  the datasets in question and/or to features of the original data
  analysis.

\end{abstract}


\onecolumn

\section{Introduction}

The power spectrum of matter is one of the most important statistics
to describe the large-scale structure of the Universe. If the
distribution of density inhomogeneities is Gaussian then the power
spectrum characterizes the distribution of matter (in a statistical
sense) completely. During the last decade considerable efforts have
been devoted to determining this function empirically from the
distribution of galaxies and clusters of galaxies. These studies have
shown that on small scales the power spectrum in real space can be
satisfactorily expressed by a power law with an index somewhere
between $-2$ and $-1.5$.  On larger scales the spectrum turns over
reaching a maximum on scales of 100 -- 150 \Mpc\ (we use a Hubble
constant of $100~h$ km~s$^{-1}$~Mpc$^{-1}$).

The exact location of the maximum, its amplitude and shape are not
well determined yet. The deep pencil-beam redshift survey of
Broadhurst \etal (1990) indicates the presence of a sharp spike at a
scale of $l=128$~\Mpc\ or wavenumber $k=2\pi/l=0.05$~\hmpc. Power
spectra of Abell-ACO clusters of galaxies (Einasto \etal 1997a,
hereafter E97a, Retzlaff \etal 1998, hereafter R98), and APM clusters
(Tadros \etal 1998, T98) also indicate a rapid turnover from a
spectrum with negative slope on galactic scales to a spectrum with
positive slope; the turnover occurs at a high amplitude on a scale
similar to the scale of the spike found by Broadhurst \etal.  But not
all power spectra obtained from galaxy surveys support this picture.
Some data show a much flatter spectrum near the maximum: the 3-D
spectrum analysis of the Las Campanas Redshift Survey (LCRS; Lin \etal
1996, hereafter LCRS3d), and the IRAS surveys discussed by Tadros and
Efstathiou (1995, hereafter TE95).

Our main goal is to determine the mean matter power spectrum using all
available data.  This will be done in three steps.  First, we derive
the mean power spectrum of galaxies that best agrees with available
observations and determine its main parameters (present Paper). By
``the mean power spectrum of galaxies'' we understand the spectrum of
a population which includes all galaxies in real space in a large
volume (fair sample).  Second, we investigate the biasing phenomenon
and develop a method to reduce the galaxy power spectrum to matter
(Einasto \etal 1999a, Paper II).  The method is based on the
assumption that the structure evolution in the Universe is due to
gravity; in this case galaxy formation is essentially a threshold
process. We find a relation between the biasing parameter and the
fraction of mass in clustered objects (galaxies).  We use numerical
simulations to follow the flow of matter from low-density to
high-density regions. In these simulations, we identify the current
epoch by comparing the $\sigma_8$ parameter of the spectrum with its
observed value.  Finally, in the third step we determine the power
spectrum of matter in the linear regime and compare it with different
model predictions (Einasto \etal 1999b, Paper III). This approach is
similar to Peacock \& Dodds (1994) but, in addition, we also determine
the primordial matter power spectrum.

The present paper is organized as follows.  In Section 2, we describe
the power spectra we shall use in our analysis. In Section 3, we
discuss why different catalogs give rise to different power spectra.
In Section 4, we derive the mean power spectrum by combining the
information from different catalogs, and determine the parameters that
define the power spectrum empirically such as its slope, amplitude,
and the shape parameter.  In Section 5 we check this spectrum for
consistency using recent determinations of the correlation function
for various galaxy and cluster samples.  Finally, we draw our main
conclusions.

\section{Power spectra from galaxy and cluster data}

In this article, our goal is to derive the mean power spectrum of all
galaxies over a wide range of scales. Our main assumption is that
there exists one single power spectrum that characterizes the
distribution of a general population of all galaxies (including giant
and dwarf galaxies, and galaxies of all morphological types) in a
large volume (fair sample of the Universe).  Real galaxy populations
are subsamples of this general galaxy population, selected in a
subvolume and in certain limited luminosity and/or morphological type
intervals of the fair sample.  Our practical task is to reduce power
spectra determined from limited galaxy populations to the fair sample.
The mean power spectrum shall be determined in or reduced to real
space.

We use the following published power spectra: the SSRS+CfA2 130 Mpc/h
volume-limited survey for $M_B < -20.3 + 5 \log h$ by da Costa \etal
(1994, hereafter dC94), the Stromlo-APM ``1--in--20'' redshift survey
of APM galaxies (Tadros and Efstathiou 1996, hereafter TE96), the
power spectrum analysis of the Las Campanas Redshift Survey (3-D
spectral analysis by LCRS3d, and 2-D analysis by Landy \etal (1996),
hereafter LCRS2d), and two IRAS surveys, the 1.2 Jy survey and the
``1--in--6'' QDOT survey (Saunders, Rowan-Robinson \& Lawrence 1992),
discussed by TE95, and by Peacock (1997, hereafter P97).  For
comparison we use power spectra of fainter galaxies: the CfA redshift
survey (Vogeley \etal 1992, Park \etal 1994), the SSRS+CfA2 101 Mpc/h
volume-limited survey for $M_B < -19.7 + 5 \log h$ by dC94; and power
spectra found by Gramann \& Einasto (1992, hereafter GE92) for
galaxies in the the Local, Coma and Perseus superclusters.  When this
study was finished we received a preprint of the power spectrum
analysis of the Durham/UKST 1 in 3 Galaxy Redshift Survey by Hoyle
\etal (1998). The power spectrum found for this survey is very similar
to the spectrum of the Stromlo-APM survey by TE96, which lies close to
our mean power spectrum $P_{HD}(k)$ (see below). This provides an
important confirmation to one of our conclusions that power spectra of
galaxy samples which cover a large volume are close to power spectra
of cluster samples, thus these samples can be considered as fair
samples of the Universe.

In addition, we use results obtained from the APM two-dimensional
galaxy distribution (Maddox \etal 1996 and references therein). This
galaxy catalog is not influenced by redshift distortions (Kaiser 1987,
Gramann \etal 1993) and, therefore, is of special value to determine
the power spectrum in real space. The problem consists in finding the
full three-dimensional power spectrum from two-dimensional data.
Recently, Peacock (P97) and Gazta\~naga \& Baugh (1998, GB98)
elaborated a procedure to use the APM Galaxy Survey.  Their results
showed a very good mutual agreement; the shapes of spectra are almost
identical, but amplitudes are slightly different. We include the mean
of these two spectra in our analysis.

The power spectra of Abell-ACO clusters (Abell 1958, Abell, Corwin and
Olowin 1989) were determined by E97a and R98; for APM clusters (Dalton
\etal 1997 and references therein) by T98. The power spectra of
clusters are similar in shape to those of galaxies (in the range of
scales not distorted by peculiar velocities) except that the cluster
power spectra are enhanced in amplitude:
$$
P_{cl}(k) = b_{cl}^{2} P_{gal}(k), 
\eqno(1)
$$
where $k$ is the wavenumber expressed in units of $h$~Mpc$^{-1}$, and
$b_{cl}$ is the bias factor of clusters relative to galaxies.  We
shall investigate the possible error of this reduction below.

Power spectra determined from galaxy and cluster surveys are 
plotted in Figure~1. We use the normalization of the power spectrum
$$
P(k)=2\pi^{2} k^{-3} \Delta^{2}(k)~,
\eqno(2)
$$
where $\Delta^{2}(k)$ is the dimensionless power spectrum (see Peacock
\& Dodds 1994).  

\begin{figure}[h]
\vspace*{10cm}
\figcaption{Observed power spectra of
galaxies and clusters of galaxies.  ACO-E and ACO-R are spectra for
Abell-ACO clusters as derived by E97 and R98; APM-T is the spectrum of
APM clusters according to TE98; APM-gal.3D and APM-gal.2D are spectra
of APM galaxies found from 3-D and 2-D data by TE96 and by P97 and
GB98, respectively; CfA2 is the spectrum of the SSRS+CfA2 130 Mpc/h
sample by dC94, LCRS is the spectrum of the LCRS according to LCRS3d;
IRAS-P and IRAS-TE are spectra of IRAS galaxies found by P97 and
TE95.}
\includegraphics{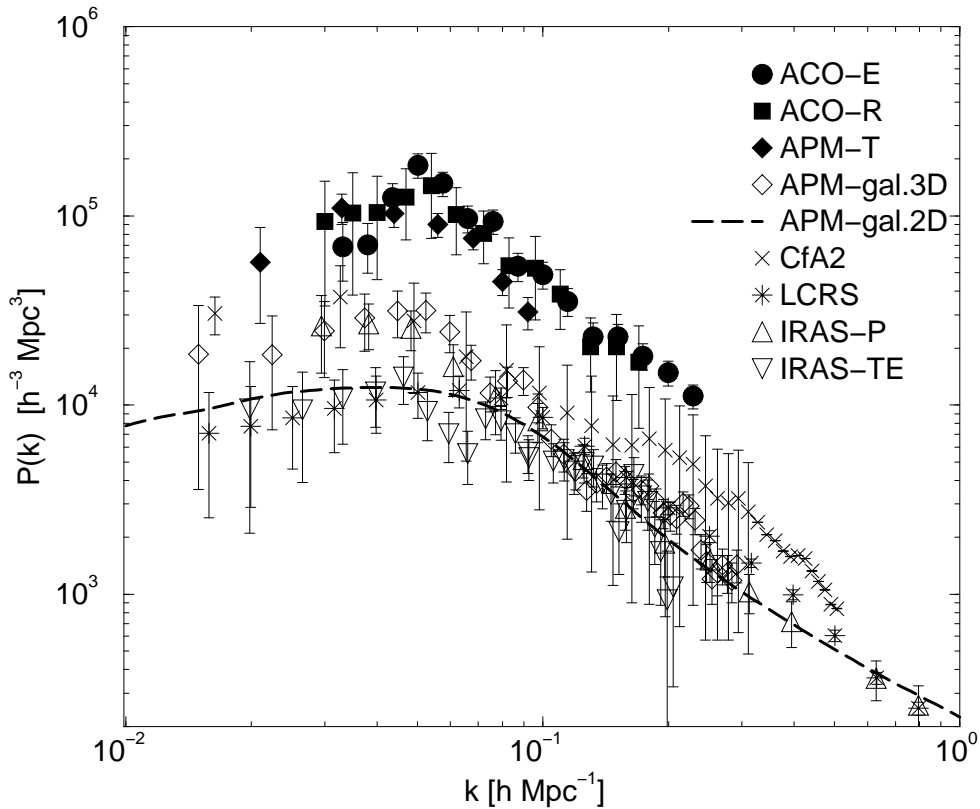} 
\label{figure1}
\end{figure}

An inspection of our Figure~1 (and a similar Figure by Vogeley (1998))
indicates that, while different spectra obtained from different
surveys agree with each other in the power law behavior on small
scales, the shape and location of the maximum is not uniquely
determined.  The spectra of clusters and APM galaxies (TE96) have a
well-determined sharp maximum, whereas IRAS and LCRS galaxies yield a
low-amplitude maximum; cluster samples indicate a rapid decline in
amplitude for $k \leq 0.05$~\hmpc\  while most galaxy samples show
a much more modest decrease. The position of the maximum for APM clusters
lies at $k\approx 0.03$~\hmpc, for Abell-ACO clusters and APM 3-D
galaxies at $k\approx 0.05$~ \hmpc, and for LCRS galaxies at $k\approx
0.06$~\hmpc.

\section{Analysis of observed power spectra}

In the previous Section we pointed out the discrepancies between the power
spectra obtained from different galaxy and cluster surveys. In this Section
we intend to clarify the source of these discrepancies and to analyze which
spectra correspond better to the actual power spectrum of all galaxies (for a
fair sample of the Universe). 

\subsection{Spectra of galaxies in high- and low-density regions}

A striking feature in Figure~1 is the difference between the power
spectrum of IRAS galaxies (as reduced by TE95) and that of Abell-ACO
clusters. It is well known that IRAS galaxies are under-represented in
high-density regions.  To understand the origin of this discrepancy
and its relation to the type of galaxy in the survey, we have
performed numerical N-body simulations.  Since spatial resolution was
crucial, we performed a 2-D analysis. We used a double-power law
initial power spectrum, which is a simple approximation of observed
spectra of galaxies and clusters of galaxies with a spike at the
maximum (Frisch \etal 1995)
$$
P(k)=\cases{A k^n,&$k\le k_0$;\cr\cr 
A k_0^n (k/k_0)^m,&$k>k_0$,} 
\eqno(3) 
$$ 
where $n$ is the power index on large scales, $m$ is the power index
on small scales and is negative, and $k_0$ is the transition
wavenumber. In our 2-D simulations we used indices $n=2$ and $m=-1$;
in the 3-D case these indices correspond to $n=1$ and $m=-2$ on large
and small scales, respectively. The turnover scale was taken to be
$1/4$ of the simulation box size.  We used a box size
of $L=512$~\Mpc.  The present epoch was identified using an rms
density dispersion of $\sigma_1=4$ on a scale of 1~\Mpc, which
corresponds to a variance of approximately $\sigma_{8}=0.9$ on a scale
of 8~\Mpc\ (Einasto \etal 1994a, hereafter E94).  We use a critical
density universe, and express densities $\varrho$ in units of the
critical (mean) density.  A top-hat smoothing over 1~\Mpc\  is used
to determine the density field. This procedure reproduces the
distribution of dark matter as accurately as possible. Dark matter
forms halos around galaxies and groups with a characteristic scale of
$\sim 1$~\Mpc\ (E94).

A density value was assigned to each particle by interpolating the
density field at the particle location. We assume that particles in
the simulation belong to different populations according to their
environment, i.e.  that galaxy samples of various environment,
morphology and luminosity can be approximated by particles in
numerical simulations chosen in certain threshold density intervals.
We shall discuss this assumption and the relation between real and
simulated galaxies in more detail in Paper II.  We call all particles
with low density values ($\varrho < \varrho_0$) {\it void particles}.
The remaining particles are clustered and form systems of various
richness; we call these clustered particles {\it galaxies}, actually
they represent dark matter in galaxies and galaxy systems. Particles
with very high density values ($\varrho \geq \varrho_{cl}$) belong to
clusters or groups, and particles with intermediate density values
($\varrho_0 \leq \varrho < \varrho_{cl}$) shall be called {\em field
galaxies}.  Here $\varrho_0$ and $\varrho_{cl}$ are threshold
densities which separate void particles from galaxies, and field
galaxies from cluster galaxies, respectively.

\begin{figure}[ht]
\vspace*{8.5cm}
\figcaption{Power spectra of simulated galaxies. The solid bold
line shows the spectrum derived for all test particles (the matter power
spectrum); various dashed and dotted lines give the power spectrum of all
galaxies, clustered galaxies in high-density regions, and galaxies in the 
intermediate density regions (simulated field galaxies). The sample
Field+.0 consist of galaxies between threshold densities $1 \leq \varrho <
5$ only; samples Field+.1 and Field+.5 contain also 10~\% and 50~\% of
galaxies selected randomly from cluster galaxies with $\varrho \geq 5$. }
\includegraphics{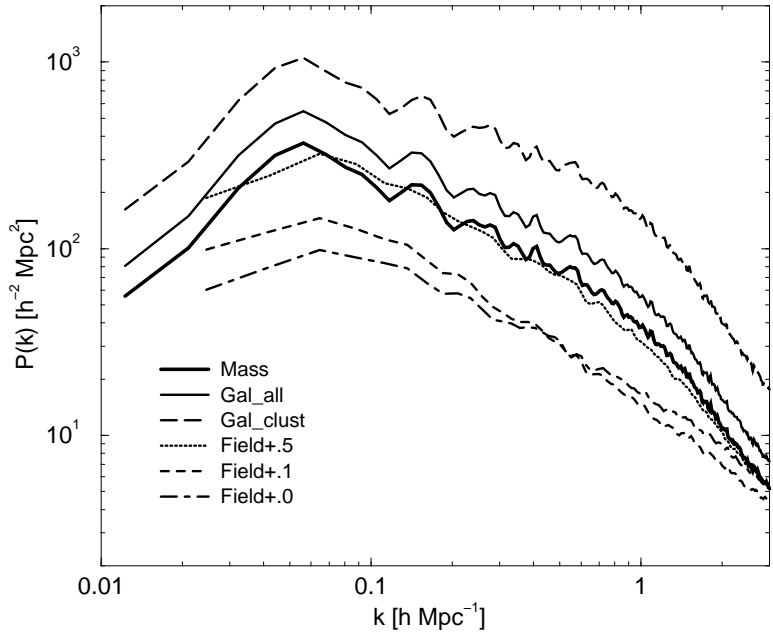} 
\label{figure2} 
\end{figure}

Figure~2 shows power spectra found for this simulation. Here we have
used threshold densities $\varrho_0=1$ and $\varrho_{cl}=5$.  In the
present paper these values serve as illustrations; their exact values
can be determined using various tests (clustering properties as
applied by E94, or hydrodynamical simulations as in Cen and Ostriker
1992, 1998).  Obviously, the power spectrum of all galaxies ($\varrho
\geq 1$) is similar to the matter  power spectrum, but has a
higher amplitude.  The power spectrum of field galaxies in the
intermediate density range ($1 \leq \varrho < 5$) has a lower
amplitude than that of the sample of all galaxies; it is even lower
than the amplitude of the matter power spectrum.  Furthermore, the
shape of the spectrum is different as well: the maximum of the
spectrum is flatter, and the power index on small scales is lower.

The population of all clustered particles (i.e. galaxies) differs from the
whole mass population in a simple way -- it does not include non-clustered
particles in low-density regions (i.e. the void population). Power spectra
are defined by the density contrast, which leads to the formula
$$
P_{m}(k) = F_c^{2} P_c(k),
\eqno(4)
$$
where $P_m(k)$ and $P_c(k)$ are power spectra of mass and clustered
particles, respectively, and $F_c$ is the fraction of matter in
clustered particles (galaxies).  The formula is exact if the
distribution of matter in low-density regions (voids) is homogeneous
(Paper II). In practice it can be used for a wide range of threshold
densities $\varrho_0$ to divide the matter into low- and high-density
regions. The form of the power spectrum of clustered particles remains
similar to that of all particles, only its amplitude has changed
according to eqn. (4). Luminous galaxies are more clustered than faint
ones; thus, varying the threshold density, one can simulate galaxy
samples of different absolute magnitude limit, and galaxies in
clusters or groups. Thus this formula can be used to reduce power
spectra of galaxies of various luminosity to the spectrum of all
galaxies, and the spectrum of all galaxies to that of all matter.  We
shall discuss possible errors of this reduction in Section 4.7 below.

The shape of the power spectrum is conserved in the case when  only
particles or galaxies in low-density regions are excluded by a certain
threshold density or luminosity limit.  If samples of particles or
galaxies are not complete in high-density (luminous galaxy) regions
then the shape of the power spectrum is not conserved (see the next
subsection).  To simulate qualitatively different catalogs we chose
galaxies in low- and medium-density regions as representatives of IRAS
galaxies, or as galaxies in poor superclusters (Einasto \etal 1997b,
1997c, hereafter E97b, E97c). In contrast to galaxies in high-density
regions, the power spectrum of galaxies in intermediate density
regions is not related to the matter power spectrum in a simple
way. Thus, it is not easy to reduce the power spectra of galaxies in
intermediate density regions to all galaxies or to all matter
distribution. But it is clear that a population, with a more
homogeneous distribution than that of all galaxies, has a power
spectrum which lies between the spectrum of a homogeneous population
(a flat spectrum of low amplitude) and the spectrum of all clustered
particles.

The main conclusions obtained from our simulation are the following.
{\em Exclusion of dark matter particles or galaxies from low-density
regions raises only the amplitude of the power spectrum in real space
without changing its shape}, whereas the {\em exclusion of galaxies
from high-density regions decreases the amplitude and changes the
shape of the power spectrum}. In the first case we can reduce data to
form a mean power spectrum characteristic for all galaxies using
eqn. (4); in the second case the shape of the power spectrum is not
conserved and the reduction of the power spectrum is complicated, thus
it is better to consider the medium-density population and its power
spectrum separately.  These results are based on the assumption that
galaxy samples can be approximated by particles in numerical
simulations chosen in certain threshold density intervals (see Section
4.7 and Paper II for a detailed analysis of this assumption).

\subsection{The influence of superclusters on IRAS data}

TE95 analyzed the power spectra of two IRAS surveys, the 1.2 Jy survey
and the ``1--in--6'' QDOT survey. They find that the spectrum of the
QDOT survey has a higher amplitude, and, if galaxies of the Hercules
supercluster are excluded, then both samples of IRAS galaxies yield a
similar power spectrum with a rather low amplitude, flat maximum.

We can consider intermediate-density (field) galaxies as representing
IRAS galaxies.  Figure~2 shows that, if we exclude all galaxies in
high-density regions from our sample (this sample is marked Field+.0 in
Figure~2), then the power spectrum of this intermediate-density
population has an amplitude near the maximum which is lower, by almost
a factor of 10, than the amplitude of the spectrum of all galaxies.
The shape of this simulated field galaxy spectrum is also different:
the maximum is much flatter.  Actually IRAS galaxies are not
completely absent in clusters, they are only under-represented. To
simulate this behavior we have formed a second simulated population of
field galaxies which includes all galaxies of the previous field
galaxy sample, and a fraction (10~\%) of cluster galaxies chosen
randomly. The corresponding mean power spectrum is shown in Figure~2
as Field+.1, it is the average of 4 sub-volumes. Its amplitude is
higher than for the previous sample, nevertheless it lies below the
simulated power spectrum of all galaxies.  By increasing the fraction
of galaxies in high-density regions to 50 ~\% (sample Field+.5) the
difference between the simulated sample of field and normal galaxies
can be reduced further. The latter spectrum is close to the spectrum of
all matter, but it has a flatter maximum.

This simple test explains qualitatively the difference observed
between the power spectrum of IRAS and normal galaxies with the
deficit of galaxies in high-density regions (in clusters and rich
superclusters) in the former sample. It is rather difficult to
calculate the correction factor to reduce such a power spectrum to the
spectrum of normal galaxies.  The correction factor depends on the
threshold density to separate field galaxies from cluster galaxies,
and on the fraction of cluster galaxies in the IRAS sample; both
parameters are not known. For this reason, in the following analysis
we consider the power spectrum of IRAS galaxies derived by TE95 as a
representative for medium-density regions in the Universe.

\subsection{Distribution of LCRS galaxies}

The power spectrum of LCRS galaxies, according to LCRS3d, is similar to
that of IRAS 1.2 Jy galaxies. It lies below the power spectrum of the
APM 3-D survey of galaxies and that of the SSRS+CfA2 sample. As we
have done before, we shall compare the distribution of LCRS galaxies
with the distribution of superclusters. Since the details of this
comparison shall be published elsewhere (Einasto \etal 1999c,
hereafter E99c), here we briefly summarize the main conclusions.

As demonstrated by Einasto \etal (1994b, 1997d, hereafter EETDA and
E97d, respectively), very rich superclusters of galaxies form a
quasi-regular network with a characteristic scale of 120~\Mpc. As
shown by E99c, the LCRS slices intersect the supercluster-void network
basically {\em in between} very rich superclusters.  Only the
Horologium-Reticulum supercluster crosses one of the southern strips
at $\alpha = 4^h$ and $cz = 20,000$ ~km/s; the $-39^{\circ}$ strip
touches also the Sculptor supercluster at $23^h~35^m$ and $cz =
33,000$ ~km/s. The rest of high-density regions observed in LCRS
coincide with clusters located in {\em poor or medium-rich}
superclusters of the catalog by E97d. Thus the most of the LCRS
strips only cross medium-density regions in the Universe.

The correlation function of clusters located in very rich
superclusters oscillates with rather high amplitude; an oscillating
correlation function corresponds to a power spectrum with a sharp
turnover near the maximum (E97b, E97c).  The correlation function of
clusters of galaxies located in poor superclusters has a lower
amplitude on large scales. The power spectrum as Fourier
transformation has a flatter maximum of lower amplitude.  These
differences between spectra resemble those described in Section
3.1. Thus the cosmography of LCRS strips suggests that near the
maximum the power spectrum of LCRS galaxies should lie {\em below} the
power spectrum of a sample in which both rich and poor superclusters
are represented. In fact, the actual power spectrum of LCRS galaxies has
a lower amplitude on large scales.  For this reason we shall use the
LCRS power spectrum as a representative of samples which include poor
and medium-rich superclusters.

On the other hand, it is possible that the low amplitude of the power
spectrum is due to the very broad window function in the Fourier
domain, caused by the narrowness of the LCRS survey in the declination
direction (see Figure~3 of LCRS3d). The 2-D power spectrum of LCRS as
derived by LCRS2d has excess power on 100~\Mpc\ scale. LCRS2d argue
that a 2-D analysis is more sensitive to the structure on large scales
than the full 3-D analysis. Then the low amplitude of the spectrum by
LCRS3d could be caused by an incomplete deprojection of the 3-D power
spectrum.  At the present stage, we can not conclude what of the two
possibilities, the different large-scale environment or problems in
data analysis, is more relevant to explain the discrepancy in the
power spectrum.  A more detailed numerical study would be required.

\subsection{The distribution of APM and ACO clusters of galaxies}

Figure~1 shows that the maxima of the power spectra found for APM and
Abell-ACO clusters of galaxies have similar amplitudes to within a
factor of 1.5.  However, for APM clusters the maximum occurs on larger
scales. In this Section we try to  clarify the reason for this
difference. A catalog of APM clusters of galaxies is now published
(Dalton \etal 1997). Thus a direct comparison of the distribution of
both cluster samples is possible.  A detailed comparison shall be
given by E99c.

The main difference between the two cluster samples is the volume they
cover.  The APM cluster sample is located only in the southern
Galactic hemisphere, and even there it covers a much smaller area on
the sky than the Abell-ACO sample.  For this reason the APM cluster
sample contains only a few very rich superclusters from the catalogs
by EETDA and E97d, (the Pisces-Cetus, Horologium-Reticulum and
Sculptor superclusters), whereas the Abell-ACO sample contains 25 very
rich superclusters (E97d). These three superclusters surround one big
void -- the Sculptor void; in contrast, the Abell-ACO catalog contains
16 voids cataloged by EETDA and surrounded by rich superclusters.
Since the Abell-ACO cluster sample covers a much larger volume in
space than the APM sample (about 4 times) we can assume that the
power spectrum found for the Abell-ACO clusters represents a larger
sample; the power spectrum of the APM cluster sample can be considered
as a local deviation. In the next Section we discuss this deviation in
more detail.

\section{Mean galaxy power spectrum}

It is our aim to determine the mean power spectrum of all galaxies in
the nearby Universe. The previous analysis has shown that
discrepancies exist between power spectra derived from various
catalogs due to differences in the spatial distribution of objects.
On the other hand, some differences may be due to differences in the
data analysis technique.  These are evident in the power spectra of
IRAS samples as discussed above, they may be present in the spectra of
LCRS galaxies. A further problem is the power spectrum reconstructed
from the 2-D distribution of APM galaxies which has a much shallower
turnover than the directly measured 3-D power spectrum (see
Figure~1). The analysis of power spectra of simulated samples has
shown that we have insufficient information to correct for all
imperfections of the data analysis techniques.

For these reasons it is not realistic to determine only one mean power
spectrum in the hope that it characterizes the distribution of all
galaxies in the whole nearby Universe. Instead, we shall determine two
mean power spectra, separately for two sets of samples (populations). In
this way we try to quantify possible differences in the distribution of
galaxies and in our ignorance of uncertainties in various data analysis
techniques. 

The first mean power spectrum was obtained from power spectra found
for cluster samples and the APM 3-D, IRAS QDOT, and SSRS+CfA2~130~Mpc
galaxy samples.  Cluster samples cover a large volume where rich
superclusters are present. We consider this population as
characteristic for high-density (HD) regions and for convenience we
call it the ``HD population''.

The second mean power spectrum was derived from spectra of the LCRS
sample, the IRAS galaxy sample as discussed by TE95, and the APM 2-D
sample. LCRS and IRAS catalogs either sample regions of the Universe
characteristic for medium-rich superclusters or samples of galaxies
where high-density regions are under-represented. We consider this
population as characteristic for medium-density (MD) regions and call
it the ``MD population''.  Let us remark that the differences between
both spectra could be due to the differences in the populations but
could also be partly an artifact of the data analysis. This could be
so in the case of the APM 2-D power spectrum.  The true power
spectrum lies probably in between, and the uncertainty range can be
understood as due to cosmic scatter (different samples) and systematic
errors.

\subsection{Galaxy power spectrum on small scales}

Observed power spectra are distorted by various effects.  Coherent
infall velocities to central regions of clusters and superclusters
increase the amplitude of the power spectrum on all scales (Kaiser
1987). Another important effect is the relative bias caused by
differences in the spatial concentration of galaxies of different
absolute magnitude and morphological type to high-density regions.
Luminous galaxies are mostly concentrated to central regions of groups
and to clusters of galaxies. These galaxies are similar to cluster
galaxies discussed above. Their power spectra are shifted to higher
amplitudes; the shift is practically scale-independent (see Figure~2,
GE92, Park \etal 1994, Peacock \& Dodds 1994, and a detailed
discussion in Paper II).  On small scales 3-D galaxy spectra are
distorted by peculiar velocities (Gramann \etal 1993) due to the
combined influence of bulk-motions and velocity dispersion of galaxies
in virialized clusters and groups.

Peacock and Dodds (1994) elaborated a technique to correct observed
spectra for these effects.  Here we shall apply a different approach:
we accept on small scales the APM galaxy power spectrum calculated
from the 2-D distribution of galaxies.  The APM 2-D data is free from
redshift distortions.  It also has some additional advantages: as it
is based on a much larger and deeper dataset, the cosmic variance is
smaller than for all presently available 3-D surveys; it contains
absolutely faint galaxies, thus we may assume that the amplitude of
this spectrum corresponds to the amplitude of the spectrum of all
galaxies.

GE92 investigated the dependence of the power spectrum on the absolute
magnitude limit $M_0$ in volume limited samples in the Local, Coma and
Perseus superclusters.  These calculations show the presence of
luminosity bias: if samples include only brighter galaxies ($M_0 \leq
-18.75 + 5\log h$) then their power spectra have a higher amplitude.
Relative bias factors (in respect to faint galaxies) for samples with
luminosity limits $M_0 = -19.75 + 5\log h$ and $M_0 = -20.25 + 5\log
h$ are 1.31 and 1.52, respectively. This luminosity bias has been
studied since then with similar results (see dC94, Park \etal 1994,
LCRS3d).  On the other hand, GE92 found that fainter galaxies have no
luminosity bias, i.e.  within sampling errors power spectra have
identical amplitudes, if the absolute magnitude limit $M_0$ lies
within the interval $-15 +5 \log h \geq M_0 > -18.75 + 5 \log h$.  In
other words, galaxy samples with sufficiently faint absolute
luminosity limits approach properties of a fair sample of the Universe
(which, by definiton, includes galaxies of all luminosities).

The APM 2-D sample is not a volume limited sample, but in the nearby
volume it includes absolutely faint galaxies, and the amplitude of the
spectrum as restored by P97 and GB98 should correspond to the sample
of all galaxies. Thus we may assume that the APM 2-D sample has small or
negligible absolute magnitude bias.

For these reasons we identify the power spectrum determined from
de-projecting the 2-D APM data on small scales ($k\geq 0.1$ \hmpc) with
the mean galaxy power spectrum. We use the mean of the power spectra by
P97 and GB98; the error estimates are practically identical, so we
accepted the errors given by P97. On larger scales we cannot apply this
power spectrum since it deviates systematically from spectra found from
3-D data.

\subsection{Galaxy power spectrum on large scales}

The main problem in calculating the mean power spectrum on larger scales
is the correction for relative bias and for redshift distortions. 

Redshift distortions are due to the contraction of superclusters (bulk
motions) and to the velocity dispersion in virialized systems
(``finger of God effect''). Bulk motions enhance the amplitude of the
power spectrum on all scales and do not change the shape of the
spectrum.  The influence of the velocity dispersion is large on small
scales where it decreases the amplitude of the spectrum.  Numerical
simulations by Gramann \etal (1993) and Suhhonenko \& Gramann (1998)
show that for realistic models the influence of velocity dispersions
to the shape of the spectrum is very small for wavenumbers $k <
0.2$~\hmpc.  Therefore, in the scales of interest we can ignore the
effect of velocity dispersion. Relative bias and redshift distortion
due to bulk motions both influence only the amplitude of the power
spectrum. Their combined effect can be determined empirically from the
comparison of amplitudes of spectra. As reference we can use the
undisturbed power spectrum of the 2-D sample near the wavenumber
$k\approx 0.1$ $h$~Mpc$^{-1}$. Figure~1 shows that the slope of power
spectra of all cluster samples is approximately the same around this
scale. This observational evidence suggests that the {\em shapes} of
3-D power spectra in this region are not distorted by redshift
effects.

Using the difference in amplitude at this wavenumber, we arrive at the
following relative bias factors: $b_{rel}=1.30$ for SSRS+CfA2 130 \Mpc\
survey (dC94), 1.12 for the APM 3-D survey (TE96), and 1.05 for the IRAS
QDOT galaxy sample (P97). A high relative bias of the SSRS+CfA2 130 \Mpc\
sample is expected as it consists of only bright galaxies. The difference
in relative bias factors of this sample from the value found by GE92 for
samples of this absolute magnitude limit (1.52) can be considered as the
uncertainty of the calibration of the amplitude of the APM 2-D spectrum as
characteristic for all galaxies. A low relative bias for IRAS galaxies is
also expected due to arguments discussed above. 

For cluster samples we find $b_{cl}=2.60$ (E97a) and $2.43$ (R98) for
Abell-ACO cluster power spectra;  and $b_{cl}=2.24$ for the APM cluster
spectrum by T98.  The spatial density of APM clusters of galaxies is
higher than that of Abell-ACO clusters; this means that APM clusters are
defined at a lower threshold density; the power spectrum of such a sample
must have a lower amplitude for reasons discussed above. 

We have determined the relative bias factors around the wavenumber
$k\approx 0.1$~\hmpc; they are somewhat smaller than found by Peacock
and Dodds (1994) for the whole scale interval. In using the whole
scale interval Peacock \& Dodds smooth the differences in the shape of
power spectra of different populations. On the other hand, it is
possible that the zero point of the amplitude as found by Peacock \&
Dodds and GE92 is closer to the true amplitude for the power spectrum
for all galaxies. If this is the case then the amplitude of our
adopted spectrum may be overestimated by a factor of up to $\approx
1.15$.  This factor characterizes the possible systematic error of the
amplitude of our mean power spectrum.

\begin{figure}[h]
\vspace*{7cm}
\figcaption{Left: Power
spectra of galaxies and clusters of galaxies normalized to the
amplitude of the 2-D APM galaxy power spectrum.  For clarity error
bars are not indicated and spectra are shown as smooth curves rather
than discrete data points.  Bold lines show spectra for clusters data,
and designations are as in Figure~1. The spectrum for APM clusters is
shifted on large scales (see text). Right: Mean power spectra derived
from galaxy spectra only ($P_{HD1}$); from galaxy and original cluster
data ($P_{HD2}$); and from galaxy and cluster data excluding APM
cluster spectrum data on scale above the maximum ($P_{HD}$). Random
errors of the mean power spectrum at $k=0.05$~\hmpc\ are 12~\%, 18~\%,
and 12~\%, respectively. Galaxy power spectra are shown for samples
CfA2~130 Mpc (dC94), IRAS-P (P97), and APM 3-D (TE96).  For comparison
the APM galaxy spectrum derived from 2-D data is also shown; it is
accepted as the mean power spectrum $P_{MD}$.  }
\includegraphics{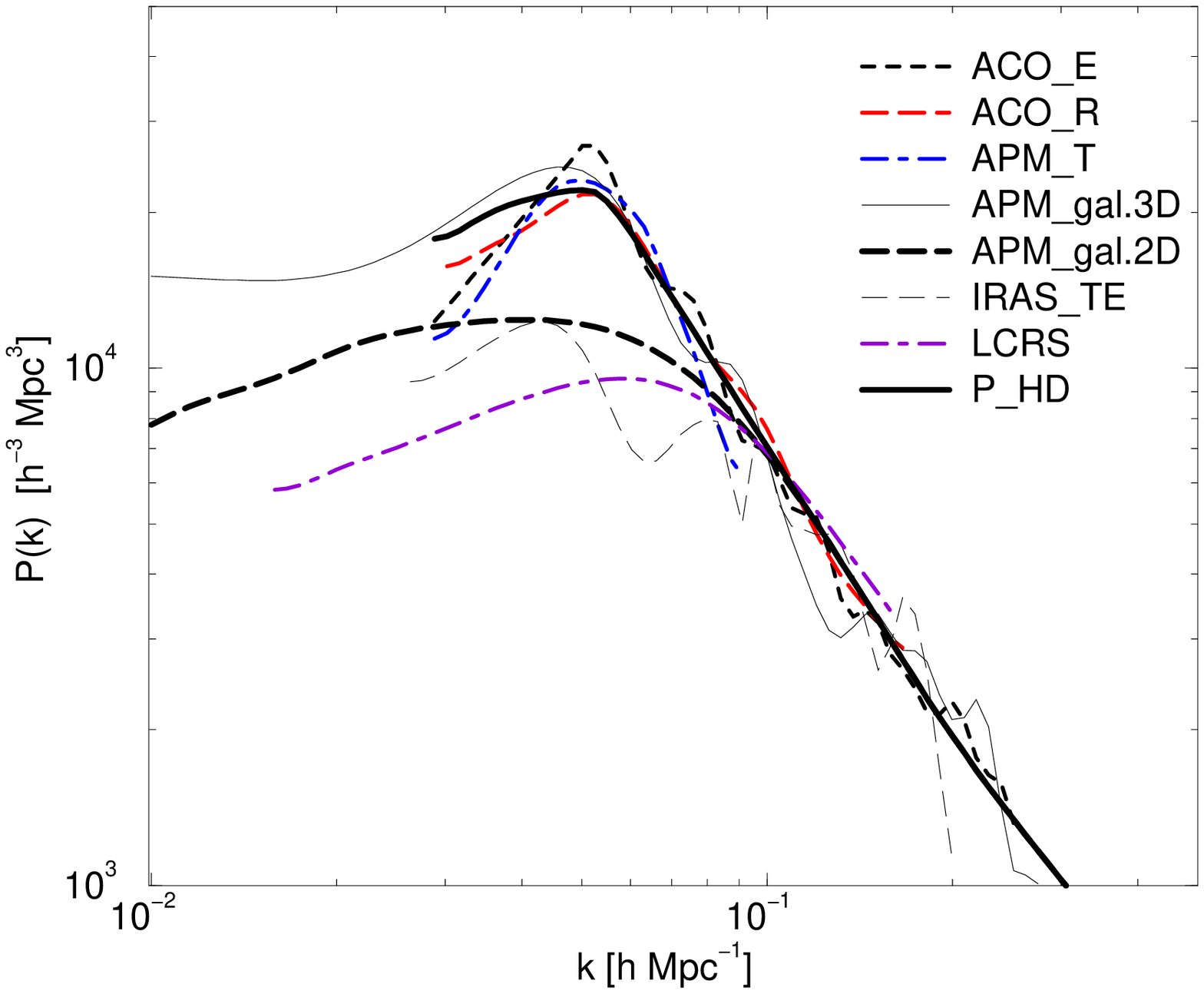} 
\includegraphics{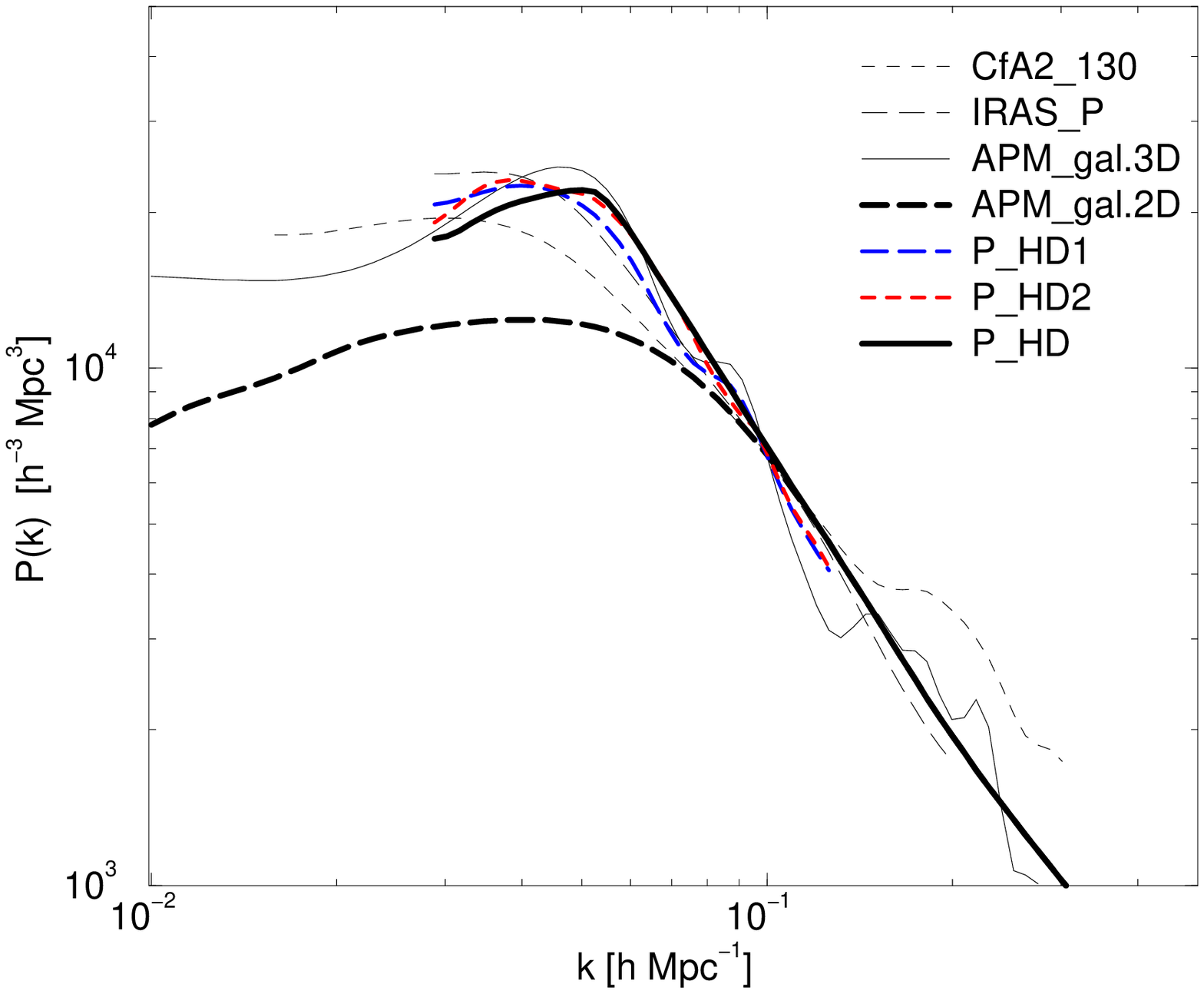}
\label{figure3}
\end{figure}

Applying these bias factors we normalized the power spectra of all
samples to the amplitude of the APM 2-D galaxy spectrum. The results
are shown in Figure~3.  The power spectra were slightly smoothed and
interpolated using a spline approximation for further analysis.

\subsection{Corrections applied to observed power spectra}

Our aim is to find the mean power spectrum of a large sample of the
Universe as accurately as possible. In order to achieve this goal we
applied a small correction to the observed spectrum in one case and
investigated the deviation from a power spectrum of a larger sample in
the other case.

The first case is the power spectrum of Abell-ACO clusters of
galaxies.  It was determined by E97a from the cluster correlation
function by Fourier transformation. This method produces an artificial
local minimum of the power spectrum near the wavenumber of $k\approx
0.035$~\hmpc. The correlation function is a Fourier transformation of
the power spectrum and vice versa if both statistics are determined in
the whole space and do not have random and systematic errors.
Actually they are determined in a limited volume and may have errors;
this particular error is due to the double-conical volume where known
clusters are located.  A correction to this effect was applied using
spectra determined for simulated clusters in N-body calculations where
the true power spectrum is known (see Section 5). Figure~2a shows that
after this correction the shapes of spectra of Abell-ACO clusters as
determined  by E97 and R98 are rather similar but not identical (see
next section for a discussion of differences between samples used by
E97 and R98).  

The second case is the APM cluster power spectrum. The mutual
separation of rich superclusters in the APM cluster sample is
anomalously large.  This is well seen in respective plots of clusters
in E99c, and is reflected in the position of the secondary peak of the
correlation function (see Figure~5 below). Rich superclusters found
for Abell-ACO clusters in the same volume have also large separations.
Large variations of separations of superclusters are common; similar
variations are also seen on void diameters (defined by surrounding
superclusters; see Table~5 of EETDA). The mean diameter of voids
defined by all Abell-ACO clusters is $91 \pm 18$~\Mpc, and the largest
deviations reach the $3\sigma$ level.  To see how the anomalously
large separation of superclusters influences the power spectrum we
shifted the APM cluster spectrum on scales above the maximum towards
shorter scales; the amount of the shift was determined from the
difference of mean separations of rich superclusters in these
catalogs.  Figure~3 (left panel) shows that after this shift the power
spectrum of APM clusters lies close to the spectrum of Abell-ACO
clusters; in other words, the difference in power spectra of Abell-ACO
and APM clusters on large scales is due to differences in the mean
separation of rich superclusters. To see the effect of this
peculiarity we find two mean power spectra, using the original data by
T98, and ignoring the APM cluster power spectrum on scales above the
maximum; results for these mean power spectra are shown in Figure~3
(right panel).

\subsection{Weights applied to determine the mean spectrum}

After the reduction of power spectra of individual samples to the
amplitude of the APM 2-D spectrum at $k\approx 0.1$~\hmpc, the
remaining differences are due to random errors, errors of relative
bias factors, cosmic scatter and/or problems in data sampling and
reduction.  In order to avoid unphysical results (negative values of
the spectrum) we assume that the error distribution of the power
spectrum is approximately lognormal.  We calculate the weighted mean
of 3-D power spectra as follows:
$$
\log P(k)={\sum \log P_i(k) w_i(k) \over \sum w_i(k)}
\eqno(5)
$$
where $P_i(k)$ is the power spectrum of sample $i$, and
$w_i(k)=w_{0i}/\sigma^2_i(k)$ is the weight of the sample at $k$; here
$w_{0i}$ is the mean weight of sample $i$,  and $\sigma_i(k)$ is the
rms error of $\log P(k)$ of sample $i$.  

For most samples we accepted original errors, i.e. we took $w_{0i}=1$.
Only the weight of the power spectrum of Abell-ACO cluster sample by
E97 was adjusted.  E97 determined the power spectrum by Fourier
transforming the correlation function, the error was calculated from
the error of the correlation function (E97b, E97c). The comparison
with other spectra suggests that the error of the power spectrum by
E97 is underestimated by a factor of $\approx 1.4$.  Correcting the
error of this sample we get the following relative weights of samples
(in units of the weight of the APM 3-D sample; in parentheses we give
the relative errors of power spectra near the maximum): 2.0 (20~\%)
and 1.0 (29~\%) for Abell-ACO cluster samples by E97a and R98,
respectively; 1.5 (23~\%) for APM clusters by T98; 1.0 (29~\%) for APM
3-D galaxy spectrum by TE96; 0.44 (48~\%) for SSRS+CfA2 galaxy sample
by dC94; and 0.88 (32~\%) for the IRAS QDOT sample by P97.  The mean
relative error of $P(k)$ averaged over all wavenumbers is 11~\%; local
deviations from this mean error are small. The volume occupied by
Abell-ACO and APM cluster samples, used by E97, R98 and T98, relates
approximately as 4:1.5:1; the number of clusters used is 869, 417 and
364, respectively for E97, R98 and T98 samples. In case of cluster
power spectra relative weights are proportional to the number of
clusters,  avoiding double-counting Abell-ACO clusters used by E97
and R98.

We calculated the mean power spectrum for three sets of weights
$w_{0i}$. In the first set we eliminated cluster samples (cluster
weights were taken $w_{0i}=0$) and used only galaxy samples with their
original weights; this mean power spectrum is denoted $P_{HD1}$ in
Figure~3b. In the second set we used the original power spectrum of
APM clusters (with weights as described above) ($P_{HD2}$ in
Figure~3b). The third set is similar to the second one, only on scales
larger than the maximum the APM cluster spectrum was ignored as it is
distorted ($P_{HD}$ in Figure~3b).  The position of the maximum of the
mean power spectrum for pure galaxy samples is $k=0.04$~\hmpc; if the
APM cluster spectrum is used then the maximum of the mean spectrum is
also at $k=0.04$~\hmpc.  The amplitude of the mean power spectrum on
the largest scales varies within a factor of $\approx 1.2$ for various
sets; the highest amplitude has the spectrum $P_{HD1}$ based on galaxy
samples only. All variations lie well within the formal error corridor
of the mean power spectrum.  We see that the mean power spectrum is
rather stable, it does not depend critically on the presence or
absence of a particular sample.

The Abell-ACO cluster sample covers the largest volume in space. The
power spectrum of APM clusters deviates on large scales with respect
to the spectrum of a fair sample of the Universe for reasons discussed
above. The mean spectrum based on galaxy spectra is obtained from
samples covering a relatively small volume. Thus we adopt the power
spectrum derived with the third set as the mean power spectrum
of all galaxies; it is shown by a solid thick line in Figure~3, and is
calculated for $k \ge 0.03$~\hmpc, the range for which data are
available for all samples used here.  For some samples, spectra are
given for even larger scales, but these data are not very accurate and
are ignored here (on largest scales the power spectrum is determined
only by a few density waves; numerical simulations show that here the
amplitude is often exaggerated, see Figure~2). This mean power
spectrum corresponds to samples which include regions of all richness
types, in particular very rich superclusters of galaxies; we denote
this power spectrum by subscript HD (for high-density).

\subsection{The mean power spectrum for medium-density regions}

On large and medium scales power spectra of LCRS galaxies (estimated
by LCRS3d) and of IRAS samples (as analyzed by TE95) are close to the
power spectrum determined from APM 2-D galaxy data.  The APM 2-D power
spectrum is, however, much smoother and has considerably smaller
random errors than LCRS and IRAS spectra.  Thus we shall use the APM
2-D power spectrum in the $k < 0.1$~\hmpc\ range as the characteristic
spectrum for the second population.

IRAS samples as reduced by TE95 represent samples of galaxies where
the number of galaxies in high-density regions is reduced. Strips of
the LCRS survey essentially intersect only poor and medium-rich
superclusters.  As noted in the introduction to this Section, we refer
to this population as ``medium-rich'' and denote the power spectrum
with subscript MD (for medium-density). This notion is for simplicity,
since it is still not clear, why the power spectrum, determined from
APM 2-D data in the scale range around the maximum, is different from
the power spectrum found from APM 3-D data. The amplitude of the power
spectrum near the maximum is determined by the spatial distribution of
very rich superclusters. It is possible that in the projection, from
three to two dimensions, part of the information on the galaxy
distribution is lost; in other words, the reduction of 2-D data to
three dimensions is imperfect.  An assessment of this and other
possibilities is beyond the scope of this paper. We continue the
discussion of differences between the power spectra of  two
populations in Section 6 using the correlation function test.

\subsection{Redshift and real space power spectra}

The power spectrum found from APM 2-D data is given in real space by
definition. We have used this spectrum in case of $P_{MD}$ on all
scales, and in case of $P_{HD}$ on medium and small scales ($k \geq
0.1$~\hmpc) -- here both spectra are identical. On larger scales the
3-D power spectra used to find the mean spectrum for the HD population
were originally determined in redshift space. We have applied the
correction due to bulk motions for all observed spectra by shifting
the observed spectra to match the APM spectrum. Thus our mean
power spectrum is reduced from redshift space to real space. Here the
assumption is that redshift distortions due to velocity dispersions
within clusters and groups can be ignored on scales of interest ($k <
0.1$~\hmpc).  Numerical simulations of CDM models with cosmological
constant show that on these scales redshift distortions due to velocity
dispersion are small (see Fig.~2 of Gramann \etal 1993).

Figure~3 shows that the mean power spectrum, $P_{HD}(k)$, has a
well-defined maximum at $k_0=0.050$~\hmpc, an approximate power law
towards smaller scales with power index $m\approx -1.9$, and a
definite decline on scales larger than that of the maximum.  The power
spectrum characteristic for medium-rich regions of the Universe,
$P_{MD}(k)$, is identical to the former spectrum over most scales, but
has a flatter maximum of lower amplitude; the maximal difference in
the amplitude is a factor of 2. Toward very large scales the
amplitudes of both power spectra definitely decrease, the decrease
being more rapid for the cluster data; the accepted mean power
spectrum $P_{HD}(k)$ is a compromise between cluster and galaxy data.

\subsection{Error analysis}

The determination of the mean power spectra of all galaxies, found
from observed power spectra of galaxies and of clusters of galaxies,
involves several steps of data reduction. Here we analyze possible
errors of these steps.

The main assumption in our analysis is that there exists a mean power
spectrum characteristic for a population which includes all galaxies
in a sufficiently large volume (fair sample). A fair sample is defined
as a sample which is characteristic for all galaxies; thus, if it
exists at all, it must have a power spectrum.  The practical questions
is: do we have any evidence that real galaxy samples approach
properties of a fair sample in this respect?  GE92 noticed that power
spectra of galaxy samples with different absolute magnitude limit are
similar in shape and have identical amplitudes, if the sample is
complete to sufficiently faint galaxies ($M_0 > -18.75 +5\log h$, see
Section 4.1).  The scatter of observed data points of power spectra
for these faint galaxy samples studied by GE92 is about 10~\% ($\pm
0.05$ in $\log P$) which can be attributed to the cosmic scatter as
the number of galaxies in samples studied for this effect was small
(from 200 to 1200). But notice that Park \etal (1994) have found a
luminosity bias for all subsamples studied ($M_0 \leq -18.5 +5\log
h$), thus the lower luminosity limit of a fair sample is not yet fixed
accurately.

The first step in the derivation of the mean power spectrum of
galaxies is the reduction of power spectra of different galaxy and
cluster samples to the spectrum of all galaxies.  Here we assume that
power spectra of different galaxy samples are similar in shape and
differ only in the overall amplitude of the spectrum. This assumption
is justified by the empirical evidence that power spectra of galaxy
and cluster samples can be brought into coincidence by adjusting
amplitudes of power spectra (see Figure~1 and Vogeley 1998). The
coincidence is, however, not exact, and we have to estimate the
corresponding error. This can be done using numerical simulations of
various galaxy samples.  

In a random Gaussian density field clusters of galaxies and samples of
galaxies of different luminosity can be considered as samples of
objects selected from the general density field by different threshold
density level.  Such an analysis is done in Paper II; it shows that a
selection by threshold density changes the amplitude of the power
spectrum (and of the correlation function), but not its shape if
samples are complete in high-density regions (see also Kaiser 1984,
GE92).  Results of numerical simulation of galaxy samples, selected at
various threshold density levels, show, that the shape of the power
spectrum in real space is conserved for galaxy samples within a
relative error of the order of 1~\%, and in the case of cluster
samples within a relative error of $\approx 5$~\%, if averaged over
the scale interval $0.01 < k \leq 1$~\hmpc (see Table~1 of Paper II),
but only 2~\%, if averaged over a scale interval $0.01 < k \leq
0.2$~\hmpc, relevant for the present study (see Figure~2b of Paper
II).  As random errors of observed power spectra values (due to the
cosmic scatter and to the Poisson noise) are of the order of 10~\% and
more (see errors shown in Figure~1), we conclude that the error
introduced to the shape of the power spectrum of all galaxies by a
shift in amplitude of power spectra of clusters and bright galaxies is
negligible.  The most serious error of the shape of the power spectrum
of clusters and bright galaxies is the enhancement of the amplitude of
the power spectrum on the largest scale. Such effect is observed in
real samples (power spectrum of the APM 3-D galaxy sample, Figure~3)
and in numerical simulations (the largest scale in model CDM6,
Figure~4).  For this reason we have ignored observed data points of
power spectra on these largest scales.

The next aspect of the data analysis is the reduction of observed
power spectra from redshift space to real space. On small scales, $k
\geq 0.1$~\hmpc, we have accepted the mean power spectrum of galaxies
on the basis of the 2-D distribution of galaxies of the APM survey,
which is given in real space by construction.  On larger scales we
have to make a distinction between redshift distortions due to the
contraction of superclusters (bulk motions) and to peculiar motion of
galaxies in virialized systems.  The first effect changes the
amplitude of the power spectrum only, the second changes also the
shape.  Numerical simulations show that on scales $k \leq 0.2$~\hmpc\
the effect of peculiar motions is small for galaxies and negligible
for clusters since cluster mean redshifts are used (Gramann \etal
1993, Suhhonenko \& Gramann 1998). This is confirmed by direct
observational data.  GE92 removed peculiar motions of galaxies in
groups and clusters by a special procedure; they found the galaxy
power spectrum to have a power index $n=-1.75$ in the range $0.08 < k
< 0.5$~\hmpc.  Within errors this coincides in the overlapping range
of scales with the power index found by GB98 for APM galaxies, and for
clusters of galaxies (E97a, R98, T97), $n=-1.9$.  We conclude that the
error of the shape of the power spectrum in real space, introduced by
this reduction procedure using original spectrum data determined in
the redshift space, is of the order $\pm 0.1$ in the power index.
 
The main error of the amplitude correction is related to the fixing of
the amplitude of the power spectrum of all galaxies in real space. We
have adopted the amplitude found by P97 and GB98 from the
reconstruction of the 3-D power spectrum from 2-D distribution of APM
galaxies.  To estimate the possible error of this normalization we
have compared relative bias factors used in this study with factors
found by GE92 for the nearby galaxy samples in the Local, Coma and
Perseus superclusters, and by P97 for other samples.  This comparison
suggests that the amplitude of our mean power spectrum can have an
error of up to 15~\%.  This normalization error is the largest
possible systematic error of the power spectrum of all galaxies.

\subsection{Parameters of the mean galaxy power spectrum}

Here we give parameters of the empirical galaxy power spectrum $P_{HD}(k)$
for samples which include rich superclusters. We recall that by
construction our power spectrum is defined in real space. The spectrum can
be specified by the following parameters:  the position of the maximum
$$
k_0=0.050 \pm 0.005~h~{\rm Mpc^{-1}}; 
\eqno(6)
$$
the amplitude of the maximum
$$
P_{HD}(k_0)=2.30 \pm 0.25 \times 10^4~h^{-3}~{\rm Mpc^3};
\eqno(7)
$$ 
and the half-width of the power spectrum at the half-maximum level
in the direction of increasing wavenumbers $k$ (Eisenstein \etal 1998)
$$ 
{\rm HWHM}=0.19 \pm 0.05~ {\rm dex}.  
\eqno(8) 
$$
(Note that the full-width cannot be estimated since the observed power
spectrum has yet to be determined for $k \lesssim 0.01~h$~Mpc$^{-1}$).

Another power spectrum parameter is the power index on intermediate
scales ($k_0 < k < 0.2$~\hmpc), which we find to be $m=-1.9 \pm 0.1$
for $P_{HD}(k)$.

The power spectrum $P_{MD}(k)$ (i.e., for samples with poor and
medium-rich superclusters) is characterized as follows: The maximum
derived from APM 2-D data occurs at wavenumber $k_0=0.040$~\hmpc, and
that derived for the LCRS at $k_0=0.063$~\hmpc; thus, the mean value
is the same as for samples including rich superclusters.  The mean
amplitude of the spectrum from APM 2-D and LCRS data is
$P_{MD}(k_0)=1.19 \pm 0.27 \times 10^4~h^{-3}~{\rm Mpc^3}$.  Formal
errors are approximately the same as for samples which include rich
superclusters. Since the power spectrum for these samples is flatter,
the half-width of the power spectrum on half-maximum level is much
larger: ${\rm HWHM}=0.42 \pm 0.10~ {\rm dex}$.  The power index on
intermediate scales is the same as for the previous case, only it
holds for a smaller range in $k$-space (see Figure~3).

Error estimates given above are determined from the mean errors of the
power spectra of individual samples used. These sampling errors are
comparable to the possible systematic error of the amplitude due to
normalization, discussed above.  The overall error of the amplitude of
the mean power spectrum of galaxies (sampling + systematic errors) is
about 20~\%.

The reason why there exist two populations of galaxies in the local
Universe with different clustering properties and power spectra is not
fully clear. The comparison of power spectra and correlation functions of
clusters of galaxies in superclusters of different richness (E97b, E97c)
hints at the possibility that we simply are dealing with regions of
different size. Very rich superclusters are rare; so, whereas larger
samples include the rarer, richer superclusters, in smaller samples only
medium and poor superclusters are present.  If this assumption is correct,
we can consider our Abell-ACO cluster sample as the closest to a fair
sample of the Universe. But this needs verification from much deeper and
larger galaxy samples. 

Table~2 contains tabulated values of the mean galaxy power spectrum
$\log P_{gal}(k)$, found for samples including rich superclusters. The
wavenumber $k$ and the spectrum are given in usual units as described
above; to allow the use of the power spectrum to calculate Fourier
integrals data are given with a small step, $\Delta \log k = 0.02$;
$\epsilon$ is the error of the logarithm of the power spectrum,
calculated from errors of individual determinations of the spectrum;
$n$ is the power index found for interval $(i-1)~-~i$ ($i$ is the row
number).  We give also the mean linear matter power spectrum, $\log
P_{lin}$, found in Papers II and III of this series, and its power
index $n$; they are also found for the power spectrum characteristic
for high-density regions.  Small waves of the linear power spectrum
are due to Doppler oscillations of the transfer function, used to
calculate theoretical power spectra applied for the extrapolation on
small scales (for details see Paper III).  The mean power spectrum for
medium-density regions, $P_{MD}(k)$, is accepted in accordance with
P97 and GB98, and is not tabulated here.

\section{The correlation function test}

In this Section we shall use various determinations of the correlation
function of galaxies and clusters of galaxies to check the consistency
of our power spectra. The correlation function and the power spectrum
are mutually related via Fourier transformation. In the absence of
random and systematic errors the observed correlation function should
be identical to the Fourier transformation of the observed spectrum
and vice versa, when known over the full range of $k$ and $r$. The
actual situation is different. Both functions are measured in a
limited range of scales; and various selection effects influence the
correlation function and the power spectrum in different ways.  For
this reason the correlation function and the power spectrum provide
two complementary methods for characterizing the large-scale structure
of the Universe.

Of special interest is the correlation function for large separations,
since it is very sensitive to the shape of the power spectrum near its
maximum.  In this region differences between various samples are more
pronounced. Also, of particular interest is the correlation function
of clusters of galaxies in rich and very rich superclusters since its
amplitude is larger  (E97b, E97c).

\subsection{Simulated correlation functions}

Let us first analyze the relation between the correlation function and
the power spectrum of matter and of clusters of galaxies in simulated
rich superclusters. Our purpose is to clarify how well high-density
regions, in particular very rich superclusters, characterize the
distribution of matter in general. For this purpose we perform 3-D
N-body simulations.  Table~1 summarizes the main parameters of our
models which differ basically in the behavior around the maximum of
the spectrum. The ``DPS'' model is the double-power law model with
sharp transition, eqn. (3), with parameters: $n=1$; $m=-1.5$; and
$k_0=0.05$~\hmpc.  The ``DPP'' model is similar to the previous one,
but its power spectrum has an additional peak of relative amplitude
$\approx 1.7$ near the maximum.  The other two models are the standard
CDM model and a flat model with cosmological constant, designated as
CDM1 and CDM2, respectively. The length of the simulation box was
taken to be $L=6 \lambda_{0}$, where $\lambda_{0}= 2\pi/k_{0}$; for
the above transition scale this gives $L=720$~\Mpc.  Calculations were
performed with a PM code with $256^3$ cells and $128^3$ particles, the
present epoch was identified by rms density fluctuations within the
simulation cell of size $l=L/256= 2.8$~\Mpc. The rms amplitude of
density perturbations of dark matter is characterized by the
$\sigma_8$ parameter; it was calculated from the power spectrum of
matter by integration (see Paper II).

\begin{table*}
\begin{center}
\caption[dummy]{Simulation parameters}
\label{tab:param}
\begin{tabular}{lcccccccc}
\\
\hline
\hline
\\
Model   & Number    & Number  & $L_{box} $ &$\Omega_{0}$&$\Omega_{\Lambda}$
& $h$ & $\sigma_8$ & $N_{cl}$\\
        & of particles&of cells & (\Mpc) & \\
\hline
\\
CDM1   & $128^{3}$ & $256^{3}$  & 720 & 1.0 & 0.0 & 0.5 & 0.57& 9350   \\
CDM2   & $128^{3}$ & $256^{3}$  & 720 & 0.2 & 0.8 & 0.5 & 0.79& 9373   \\
DPS    & $128^{3}$ & $256^{3}$  & 720 & 1.0 & 0.0 & 0.5 & 0.88& 9339   \\
DPP    & $128^{3}$ & $256^{3}$  & 720 & 1.0 & 0.0 & 0.5 & 0.79& 9445   \\

CDM6   & $240^{3}$ & $720^{3}$  & 720 & 1.0 & 0.0 & 0.5 & 0.46& 9329   \\
\\
\hline
\label{tab:prop}
\end{tabular}
\end{center}
\end{table*}

Clusters of galaxies were identified with a tree code which picks up
high-density clumps of particles. Parameters of the neighborhood
search were chosen so as to obtain a total of $2\times 10^4$ clusters
in each model. From this initial cluster sample a subsample was
selected choosing the $N_{cl}\approx 9300$ most massive candidates
(the mass is determined from the number of particles included).  This
corresponds to the spatial density of Abell-ACO clusters of $n =
2.5\times 10^{-5}~h^{-3}~Mpc^3$ in this volume, taking into account
the selection function of real clusters both in galactic latitude and
distance (see E97b and E97d for details).  

These simulations have a low spatial resolution. To check how
sensitive the clustering properties are to the chosen resolution, we
calculated a new variant (CDM6, with much higher resolution) for the
standard CDM model. This model was calculated with a different
algorithm and different realization of the initial density field; it
has a lower $\sigma_8$ value, thus its amplitude is lower.  The lower
mass limit of clusters is also different since the rate of cluster
formation depends on the amplitude of the power spectrum on large
scales.

\begin{figure} 
\vspace*{16.5cm}
\figcaption[]{Spectra and correlation functions of N-body models.
Upper panel: matter spectra; middle panel: correlation functions
calculated from matter spectra by Fourier transformation and enhanced in
amplitude to simulate cluster correlation functions; lower panel:
actual correlation functions of model clusters in rich superclusters. 
Models are designated as in Table~1.}
\includegraphics{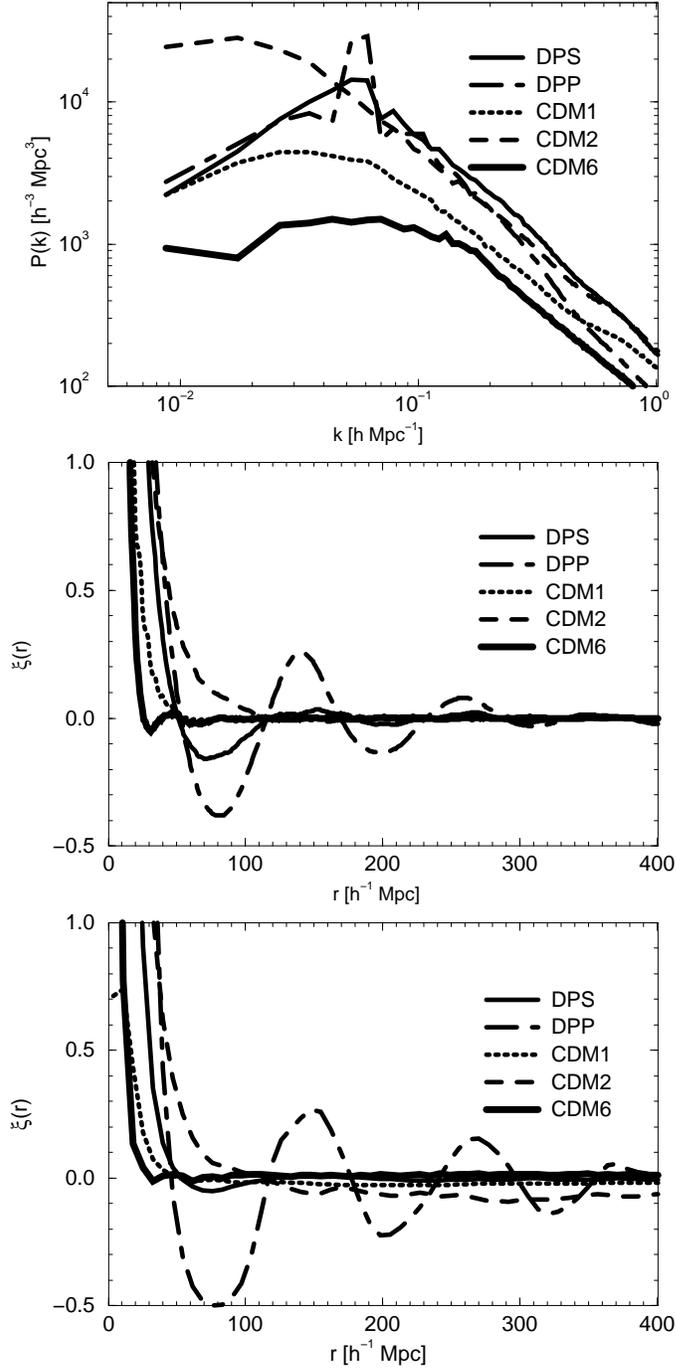}
\label{fig:figure4}
\end{figure}

Matter power spectra derived from all test particles are shown in the
upper panel of Figure~4. As indicated by the $\sigma_8$ value, the model
CDM6 has a spectrum of lower amplitude.  In the present context the
absolute normalization of the amplitude of the power spectrum is of minor
importance as we compare cluster correlation functions for the same model,
calculated directly from simulated cluster samples and indirectly from the
Fourier transformation of the matter power spectrum. On large scales the
power spectrum of the CDM6 model is rising, which is probably a numerical
artifact (similar deviations occur if we extract from the whole simulation
box a smaller sub-volume and find the power spectrum there). 

These spectra were used to calculate the correlation function of matter
using the Fourier transformation. As demonstrated already by Kaiser
(1984), the correlation function of clusters of galaxies has a higher
amplitude than the correlation function of galaxies. The same is true
for power spectra. The analysis by E97b and E97c has shown that the
amplitude of the correlation functions (and of the power spectra) of
clusters in very rich superclusters is larger than that for galaxies, by a
factor of about 50, which corresponds to a relative bias factor of 7.
Thus, in order to find the expected correlation function of clusters in
very rich superclusters, we multiply the matter correlation functions of
our models by this factor (here we ignore the difference between the
matter and galaxy power spectra and correlation functions as this
difference is small, see below). The resulting expected correlation
functions are plotted in the middle panel of Figure~4. 

Using model cluster catalogs we constructed model superclusters applying
a procedure identical to the determination of real superclusters (EETDA,
E97d).  The correlation functions of clusters of galaxies located in very
rich superclusters with 8 or more members are shown in the lower panel of
Figure~4. We see that these correlation functions are rather similar to
expected correlation functions calculated from the distribution of all
particles by Fourier transforming the matter power spectra.

We conclude that simulated superclusters, in particular
very rich superclusters, can be used to describe the matter
distribution of the whole model. A check with high-resolution
simulation CDM6 confirms results obtained with medium resolution. In
all CDM models the distribution of superclusters is much less regular
than in DPS and especially in DPP model.  These geometric properties
are well expressed through the correlation functions and power spectra
of respective models (for the geometric interpretation of correlation
functions see E97b, E97c). Thus we may hope that the correlation
function of real rich superclusters can be used to test the
distribution of the whole matter in the Universe. In particular, we
note that the correlation function is very sensitive to the shape of
the power spectrum around the maximum. The correlation function is
oscillating only in the case when the power spectrum has a sharp peak
at the maximum, otherwise it approaches the zero level on large
scales.

A similar conclusion has been obtained by Suhhonenko and Gramann (1998) 
using high-resolution N-body simulations and analytical calculations based 
on the Press-Schechter (1974) algorithm.

\subsection{Observed correlation functions}

Now we compare the correlation function for clusters in very rich
superclusters as derived from Abell-ACO clusters with that from APM
clusters (E97b, E99c). As can be seen in the left panel of Figure~5,
our data show clearly that the correlation function of clusters in
very rich superclusters has a well-defined secondary maximum for both
cluster samples.  The secondary maximum of the cluster correlation
function is due to the correlation of clusters across large voids.
The mutual separation of very rich superclusters of the APM cluster
sample is rather large. This separation determines the location of the
secondary maximum of the correlation function, which is 185~\Mpc,
determined both from the APM cluster sample, and from the Abell-ACO
cluster sample in the same volume. The secondary maximum of the
correlation function for the whole Abell-ACO cluster sample is located
at a separation of 140~\Mpc.  The amplitude of the secondary maximum
for APM clusters is exaggerated since the sample is small.  In a small
sample the number of rich superclusters is small (APM sample contains
only 3 very rich superclusters); the secondary maximum of the
correlation function is given by the mutual separation of these few
superclusters. In a large sample there are many rich superclusters,
all of them have different separations. Local secondary maxima of the
correlation function (found in small subvolumes) cancel each other
partly out, and the secondary maximum of the mean correlation function
calculated for the whole volume has a much lower amplitude than maxima
in smaller subvolumes.

\begin{figure}[ht]
\vspace*{7cm}
\figcaption{ The correlation functions of galaxies and clusters of
 galaxies.  The left panel gives these functions for clusters of
 galaxies located in very rich superclusters of Abell-ACO and APM
 samples.  The right panel shows the correlation functions of galaxies
 in the APM 3-D ``1--in--20'' sample (Loveday \etal 1995) and LCRS
 sample (Tucker \etal 1997); Lov-me, Lov-ml, Lov-bl denote
 medium-bright early type, medium-bright late type, and bright late
 type galaxies from Figure~6 of Loveday \etal. For comparison we show the
 correlation functions calculated by Fourier transformation of power
 spectra, $P_{HD}(k)$ (bold solid line), and $P_{MD}(k)$ (bold dashed
 line). All galaxy correlation functions are enhanced in amplitude to
 match the correlation functions for clusters in rich superclusters.
 }
\includegraphics{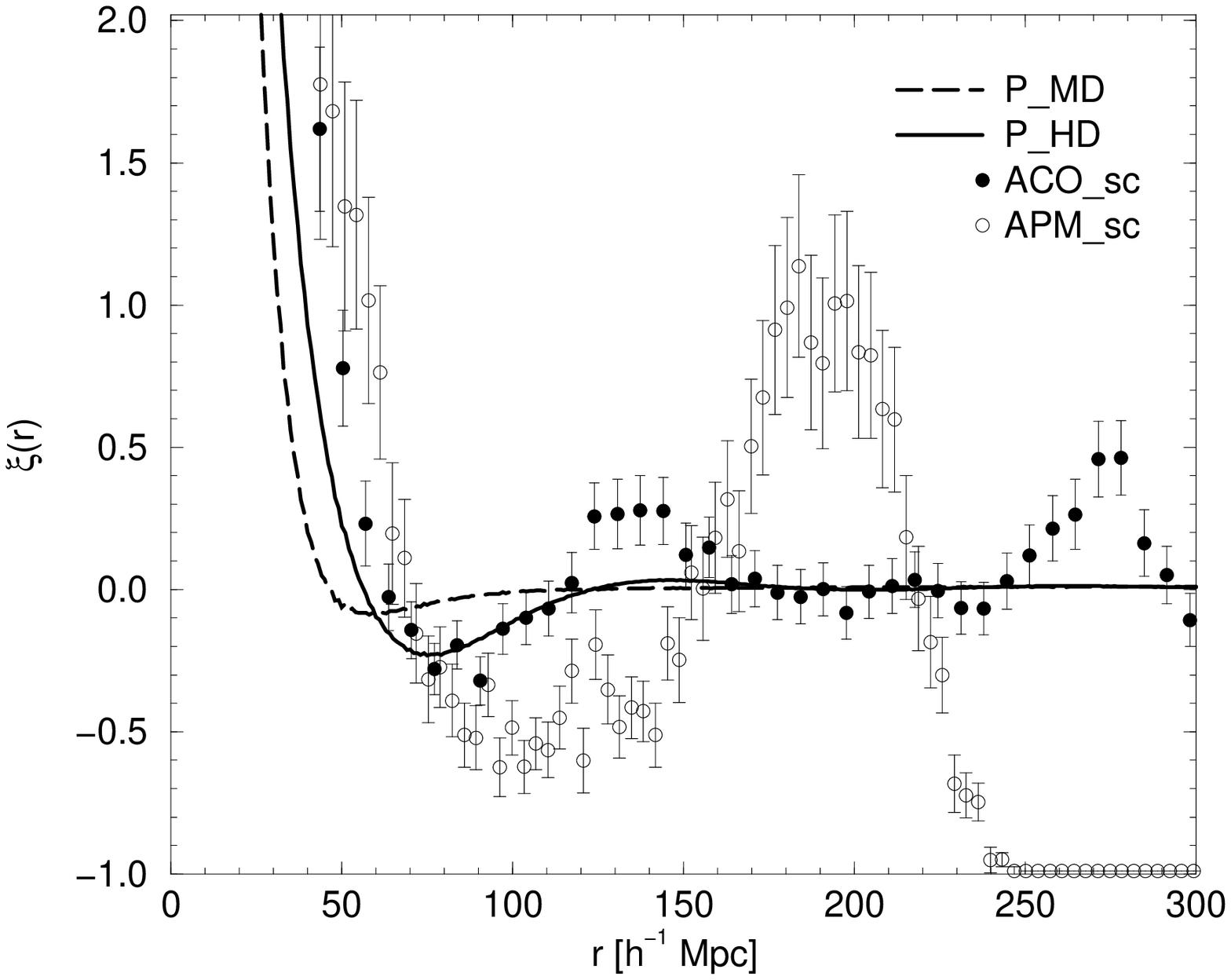} 
\includegraphics{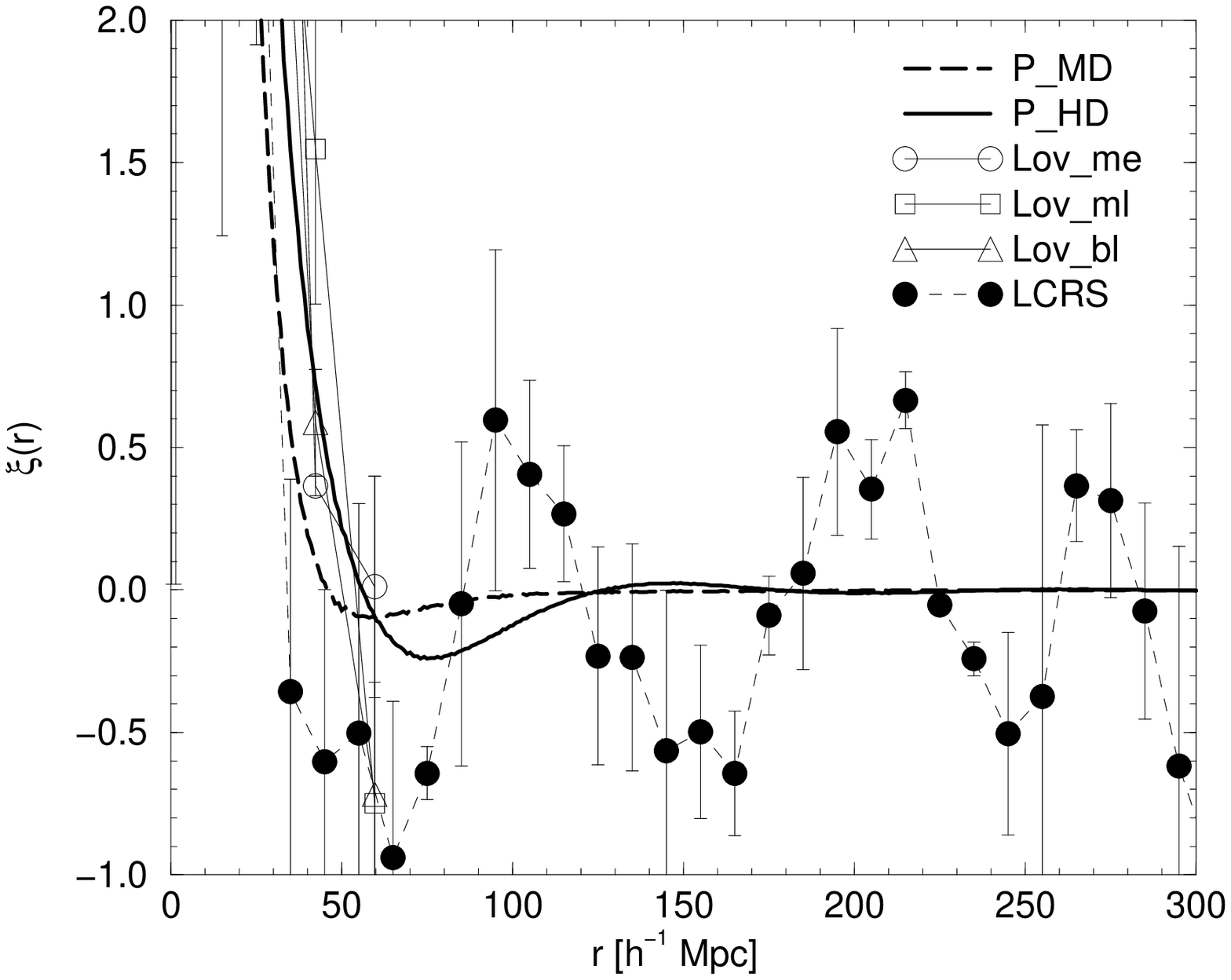}
\label{figure5}
\end{figure}

The location of the secondary maximum of the correlation function is
closely connected with the position of the maximum of the power
spectrum (E97b, E97c).  As both the spectrum and the correlation
function of the APM cluster sample are determined from a much smaller
volume than those for the Abell-ACO sample, we have to conclude that the
differences in the power spectrum and the correlation function derived
for these two  cluster samples are due to cosmic fluctuations.

In Figure~5 we also give the correlation functions calculated from the
observed power spectrum of matter, $P_{HD}(k)$, as adopted in the
previous Section, as well as from the spectrum $P_{MD}(k)$ which has a
flat maximum.  These functions were also matched in amplitude to that
of the cluster correlation function in rich superclusters. The
correlation function calculated from the power spectrum $P_{HD}(k)$
has a zero crossing at 60 \Mpc, not far from the zero crossing of the
correlation function of clusters in Abell-ACO (60~\Mpc) and APM
(70~\Mpc) rich superclusters. On smaller separations its amplitude is
lower than the observed cluster correlation amplitude. The correlation
function derived from the power spectrum $P_{MD}(k)$ has a much lower
zero crossing (about 45 \Mpc) and amplitude on small scales. 

Cluster correlation functions can be compared with galaxy correlation
functions derived by Loveday \etal (1995) for the APM ``1--in--20''
redshift survey and for LCRS galaxies (Tucker \etal 1997), as indicated in
the right panel of Figure~5.  Unfortunately, Loveday et al. calculated the
galaxy correlation functions only for $r \leq 60$~\Mpc, so that the
presence or absence of a secondary maximum of the galaxy correlation
function cannot be established. For the present comparison the position of
zero crossing is important, as it does not depend on the relative bias
factor. For various subsamples of the APM galaxy sample the zero crossing
lies between 50 and 55 \Mpc, a bit less than for the correlation function
derived from the power spectrum $P_{HD}(k)$ (60~\Mpc).  The correlation
function of galaxies in the LCRS galaxies (Tucker \etal 1997) has a zero
crossing (32~\Mpc) not far from that of the correlation function derived
from the power spectrum $P_{MD}(k)$ (45~\Mpc). Note also that the
correlation function of LCRS galaxies is oscillating as the cluster
correlation function. The period of oscillations is approximately
100~\Mpc, as expected from the position of the maximum of the power
spectrum of LCRS galaxies, and from the separation of rich galaxy
filaments in LCRS slices.  We see that correlation functions are in
satisfactory agreement with power spectra. 

We conclude that the correlation function test reinforces our findings
on the basis of the power spectra. There exist differences in
clustering properties of populations in the nearby Universe. One
population is characteristic for rich superclusters, and the other for
poorer ones. The former population has a power spectrum with a sharp
peak and a correlation function with zero crossing near 60~\Mpc, the
latter population has a flatter power spectrum and a zero crossing of
the correlation function near 40 \Mpc.  On the other hand, all power
spectra of samples of the MD population have one or other problem --
either with limitations inherent to the dataset itself or with the
appropriateness of the data analysis.  Thus, the differences between
our adopted mean power spectra of high and medium density regions
could be partly attributed to these problems.

\section{Conclusions}

In this paper our goal is to determine the mean power spectrum of
galaxies in real space in a large representative volume. Our principal
assumption is that a fair sample has a power spectrum, i.e. a mean
spectrum characteristic for a population which includes all galaxies
including faint dwarf galaxies.

We have used observed power spectra determined from deep galaxy and
cluster samples.  In analyzing these samples we have found two
distinct populations.  The first population is characteristic of
volumes containing superclusters of a wide variety of richness
classes; samples which can be identified with this first population
include the Abell-ACO and the APM cluster catalogs, the 3-D redshift
survey of APM galaxies, and the SSRS+CfA2 130 Mpc galaxy catalog.  The
second population is characteristic of volumes which -- due to smaller
volume and/or unfortunate survey geometry -- lack the rarer, richer
superclusters; this ``medium-rich'' population is represented by the
LCRS (whose slice geometry seems to be anti-correlated with the
positions of the richest superclusters) and the IRAS galaxy sample
(since IRAS galaxies tend to avoid high-density environments).

We have found mean power spectra for these populations.  Mean power
spectra were reduced to correct for redshift distortions due to bulk
motions.  On medium and small scales the power spectra of the two
populations are identical. This is because on these scales, the mean
power spectrum is adopted from 3-dimensional reconstruction of the
2-dimensional distribution of APM galaxies, which is free of redshift
distortions and, due to the very large size of the sample, has a small
cosmic scatter.

On large scales, the mean power spectrum which we adopted for the
high-density population, $P_{HD}(k)$, was derived from observed power
spectra in redshift space for cluster and galaxy samples listed above.
Redshift space spectra are reduced to real space and corrected for the
relative bias. On these scales redshift distortions due to peculiar
motions of galaxies in clusters and groups are small and can be
ignored.  The mean power spectrum has a fairly sharp maximum at
$k=0.05 \pm 0.01$~\hmpc, and a half-width half-maximum of ${\rm
HWHM}=0.19 \pm 0.05~ {\rm dex}$; on smaller scales, it exhibits an
almost exact power law of index $n=-1.9$. Clusters of galaxies have
the largest weight in the determination of this power spectrum.

The power spectrum we find for medium-density regions, $P_{MD}(k)$,
has a flatter maximum with ${\rm HWHM}=0.42 \pm 0.10~ {\rm dex}$.

We have argued that clusters of galaxies adequately trace the true
mass distribution in the Universe, so that the cluster data based
power spectrum of galaxies can probably be considered as the power
spectrum of a fair sample of the Universe.

\noindent{\it Acknowledgements:} We thank E. Gazta\~naga, H. Tadros, and
M.  Vogeley for supplying the Tartu group with the numerical values of
galaxy and cluster power spectra, and M. Gramann and A.  Szalay for
discussion. We thank the referee, M. Vogeley, for constructive criticism. 
This work was supported by the Estonian Science Foundation (grant 2625),
International Science Foundation (grant LLF100). JE and AS were supported
by the Deutsche Forschungsgemeinschaft in Potsdam; AS was partially
supported by the Russian Foundation for Basic Research under Grant
96-02-17591; RC was supported by grants NAG5-2759, AST9318185; HA
benefitted from financial support by CONACYT (Mexico; C\'atedra
Patrimonial, ref 950093). 


\scriptsize
\twocolumn
\begin{table}
\begin{center}
\caption{Power spectra}
\begin{tabular}{rrrrrr}
\\
\hline
\hline
\\
$\log k$ & $\log P_{gal}$ & $\epsilon$ & $n$ & $\log P_{lin}$ & $n$ \\
\\
\hline
\\
\\
\\
 -1.54 &     4.2350 &    .0891 &      .75 &     3.9850 &      .75 \\  
 -1.52 &     4.2425 &    .0853 &      .37 &     3.9925 &      .37 \\  
 -1.50 &     4.2553 &    .0832 &      .64 &     4.0053 &      .64 \\  
\\
 -1.48 &     4.2710 &    .0806 &      .79 &     4.0210 &      .79 \\  
 -1.46 &     4.2867 &    .0773 &      .78 &     4.0367 &      .78 \\  
 -1.44 &     4.3000 &    .0732 &      .66 &     4.0500 &      .66 \\  
 -1.42 &     4.3104 &    .0685 &      .52 &     4.0604 &      .52 \\  
 -1.40 &     4.3189 &    .0637 &      .43 &     4.0689 &      .43 \\  
\\
 -1.38 &     4.3270 &    .0593 &      .40 &     4.0770 &      .40 \\  
 -1.36 &     4.3356 &    .0554 &      .43 &     4.0856 &      .43 \\  
 -1.34 &     4.3447 &    .0521 &      .46 &     4.0947 &      .46 \\  
 -1.32 &     4.3534 &    .0498 &      .44 &     4.1034 &      .44 \\  
 -1.30 &     4.3614 &    .0482 &      .40 &     4.1114 &      .40 \\  
\\
 -1.28 &     4.3549 &    .0412 &     -.33 &     4.1049 &     -.33 \\  
 -1.26 &     4.3332 &    .0401 &    -1.09 &     4.0720 &     -.95 \\  
 -1.24 &     4.2996 &    .0393 &    -1.68 &     4.0437 &    -1.41 \\  
 -1.22 &     4.2604 &    .0384 &    -1.96 &     4.0099 &    -1.69 \\  
 -1.20 &     4.2229 &    .0375 &    -1.87 &     3.9725 &    -1.87 \\  
\\
 -1.18 &     4.1856 &    .0366 &    -1.87 &     3.9351 &    -1.87 \\  
 -1.16 &     4.1481 &    .0361 &    -1.87 &     3.8978 &    -1.87 \\  
 -1.14 &     4.1107 &    .0358 &    -1.87 &     3.8604 &    -1.87 \\  
 -1.12 &     4.0733 &    .0360 &    -1.87 &     3.8230 &    -1.87 \\  
 -1.10 &     4.0359 &    .0366 &    -1.87 &     3.7856 &    -1.87 \\  
\\
 -1.08 &     3.9985 &    .0375 &    -1.87 &     3.7483 &    -1.86 \\  
 -1.06 &     3.9611 &    .0383 &    -1.87 &     3.7109 &    -1.87 \\  
 -1.04 &     3.9237 &    .0384 &    -1.87 &     3.6735 &    -1.87 \\  
 -1.02 &     3.8863 &    .0426 &    -1.87 &     3.6361 &    -1.87 \\  
 -1.00 &     3.8489 &    .0421 &    -1.87 &     3.5988 &    -1.87 \\  
\\
  -.98 &     3.8115 &    .0422 &    -1.87 &     3.5614 &    -1.87 \\  
  -.96 &     3.7741 &    .0421 &    -1.87 &     3.5240 &    -1.87 \\  
  -.94 &     3.7367 &    .0416 &    -1.87 &     3.4866 &    -1.87 \\  
  -.92 &     3.6993 &    .0413 &    -1.87 &     3.4493 &    -1.86 \\  
  -.90 &     3.6632 &    .0815 &    -1.81 &     3.4132 &    -1.81 \\  
\\
  -.88 &     3.6262 &    .0785 &    -1.85 &     3.3762 &    -1.85 \\  
  -.86 &     3.5887 &    .0752 &    -1.87 &     3.3387 &    -1.87 \\  
  -.84 &     3.5509 &    .0716 &    -1.89 &     3.3009 &    -1.89 \\  
  -.82 &     3.5128 &    .0678 &    -1.90 &     3.2628 &    -1.91 \\  
  -.80 &     3.4748 &    .0640 &    -1.90 &     3.2248 &    -1.90 \\  
\\
  -.78 &     3.4369 &    .0601 &    -1.89 &     3.1711 &    -2.69 \\  
  -.76 &     3.3994 &    .0562 &    -1.87 &     3.1221 &    -2.45 \\  
  -.74 &     3.3625 &    .0525 &    -1.85 &     3.0810 &    -2.05 \\  
  -.72 &     3.3263 &    .0489 &    -1.81 &     3.0461 &    -1.75 \\  
  -.70 &     3.2909 &    .0456 &    -1.77 &     3.0116 &    -1.72 \\  
\\
\hline
\end{tabular}
\end{center}
\end{table}

\begin{table}
\begin{center}
\vskip0.75cm
\begin{tabular}{rrrrrr}
\\
\hline
\hline
\\
$\log k$ & $\log P_{gal}$ & $\epsilon$ & $n$ & $\log P_{lin}$ & $n$ \\
\\
\hline
\\
  -.68 &     3.2566 &    .0425 &    -1.72 &     2.9709 &    -2.03 \\  
  -.66 &     3.2232 &    .0397 &    -1.67 &     2.9217 &    -2.46 \\  
  -.64 &     3.1906 &    .0373 &    -1.63 &     2.8675 &    -2.71 \\  
  -.62 &     3.1588 &    .0351 &    -1.59 &     2.8160 &    -2.58 \\  
  -.60 &     3.1277 &    .0333 &    -1.56 &     2.7720 &    -2.20 \\  
\\
  -.58 &     3.0972 &    .0320 &    -1.53 &     2.7328 &    -1.96 \\  
  -.56 &     3.0672 &    .0310 &    -1.50 &     2.6904 &    -2.12 \\  
  -.54 &     3.0376 &    .0304 &    -1.48 &     2.6412 &    -2.46 \\  
  -.52 &     3.0085 &    .0301 &    -1.45 &     2.5893 &    -2.59 \\  
  -.50 &     2.9798 &    .0302 &    -1.44 &     2.5415 &    -2.39 \\  
\\
  -.48 &     2.9515 &    .0305 &    -1.42 &     2.4977 &    -2.19 \\  
  -.46 &     2.9234 &    .0311 &    -1.40 &     2.4519 &    -2.29 \\  
  -.44 &     2.8956 &    .0319 &    -1.39 &     2.4025 &    -2.47 \\  
  -.42 &     2.8681 &    .0327 &    -1.38 &     2.3540 &    -2.43 \\  
  -.40 &     2.8407 &    .0337 &    -1.37 &     2.3060 &    -2.40 \\  
\\
  -.38 &     2.8136 &    .0348 &    -1.35 &     2.2578 &    -2.41 \\  
  -.36 &     2.7868 &    .0359 &    -1.34 &     2.2096 &    -2.41 \\  
  -.34 &     2.7602 &    .0370 &    -1.33 &     2.1615 &    -2.41 \\  
  -.32 &     2.7339 &    .0381 &    -1.31 &     2.1132 &    -2.41 \\  
  -.30 &     2.7079 &    .0392 &    -1.30 &     2.0645 &    -2.43 \\  
\\
  -.28 &     2.6822 &    .0402 &    -1.28 &     2.0156 &    -2.45 \\  
  -.26 &     2.6569 &    .0412 &    -1.27 &     1.9667 &    -2.45 \\  
  -.24 &     2.6319 &    .0421 &    -1.25 &     1.9178 &    -2.44 \\  
  -.22 &     2.6072 &    .0429 &    -1.23 &     1.8689 &    -2.45 \\  
  -.20 &     2.5829 &    .0437 &    -1.22 &     1.8199 &    -2.45 \\  
\\
  -.18 &     2.5589 &    .0444 &    -1.20 &     1.7708 &    -2.45 \\  
  -.16 &     2.5352 &    .0450 &    -1.18 &     1.7216 &    -2.46 \\  
  -.14 &     2.5119 &    .0456 &    -1.17 &     1.6724 &    -2.46 \\  
  -.12 &     2.4889 &    .0462 &    -1.15 &     1.6231 &    -2.46 \\  
  -.10 &     2.4660 &    .0467 &    -1.14 &     1.5739 &    -2.46 \\  
\\
  -.08 &     2.4432 &    .0471 &    -1.14 &     1.5245 &    -2.47 \\  
  -.06 &     2.4203 &    .0474 &    -1.14 &     1.4752 &    -2.47 \\  
  -.04 &     2.3974 &    .0476 &    -1.15 &     1.4258 &    -2.47 \\  
  -.02 &     2.3742 &    .0478 &    -1.16 &     1.3764 &    -2.47 \\  
   .00 &     2.3507 &    .0478 &    -1.17 &     1.3270 &    -2.47 \\  
\\
   .02 &     2.3268 &    .0477 &    -1.19 &     1.2776 &    -2.47 \\  
   .04 &     2.3027 &    .0475 &    -1.21 &     1.2283 &    -2.47 \\  
   .06 &     2.2783 &    .0472 &    -1.22 &     1.1789 &    -2.47 \\  
   .08 &     2.2536 &    .0468 &    -1.23 &     1.1295 &    -2.47 \\  
   .10 &     2.2287 &    .0463 &    -1.25 &     1.0802 &    -2.47 \\  
\\
   .12 &     2.2035 &    .0458 &    -1.26 &     1.0308 &    -2.47 \\  
   .14 &     2.1781 &    .0451 &    -1.27 &      .9815 &    -2.47 \\  
   .16 &     2.1524 &    .0443 &    -1.28 &      .9322 &    -2.46 \\  
   .18 &     2.1264 &    .0435 &    -1.30 &      .8830 &    -2.46 \\  
   .20 &     2.1001 &    .0426 &    -1.31 &      .8337 &    -2.46 \\  
\\
\hline
\end{tabular}
\end{center}
\end{table}

\begin{table}
\begin{center}
\begin{tabular}{rrrrrr}
\\
\hline
\hline
\\
$\log k$ & $\log P_{gal}$ & $\epsilon$ & $n$ & $\log P_{lin}$ & $n$ \\
\\
\hline
\\
   .22 &     2.0735 &    .0417 &    -1.33 &      .7845 &    -2.46 \\  
   .24 &     2.0464 &    .0407 &    -1.35 &      .7352 &    -2.46 \\  
   .26 &     2.0189 &    .0397 &    -1.38 &      .6860 &    -2.46 \\  
   .28 &     1.9909 &    .0387 &    -1.40 &      .6368 &    -2.46 \\  
   .30 &     1.9625 &    .0377 &    -1.42 &      .5876 &    -2.46 \\  
\\
   .32 &     1.9339 &    .0367 &    -1.43 &      .5384 &    -2.46 \\  
   .34 &     1.9050 &    .0357 &    -1.44 &      .4891 &    -2.46 \\  
   .36 &     1.8760 &    .0347 &    -1.45 &      .4399 &    -2.46 \\  
   .38 &     1.8470 &    .0338 &    -1.45 &      .3907 &    -2.46 \\  
   .40 &     1.8180 &    .0329 &    -1.45 &      .3414 &    -2.46 \\  
\\
   .42 &     1.7891 &    .0320 &    -1.45 &      .2922 &    -2.46 \\  
   .44 &     1.7603 &    .0312 &    -1.44 &      .2430 &    -2.46 \\  
   .46 &     1.7316 &    .0304 &    -1.43 &      .1938 &    -2.46 \\  
   .48 &     1.7031 &    .0297 &    -1.43 &      .1446 &    -2.46 \\  
   .50 &     1.6747 &    .0289 &    -1.42 &      .0954 &    -2.46 \\  
\\
   .52 &     1.6465 &    .0283 &    -1.41 &      .0462 &    -2.46 \\  
   .54 &     1.6184 &    .0276 &    -1.40 &     -.0030 &    -2.46 \\  
   .56 &     1.5904 &    .0269 &    -1.40 &     -.0522 &    -2.46 \\  
   .58 &     1.5625 &    .0263 &    -1.39 &     -.1014 &    -2.46 \\  
   .60 &     1.5346 &    .0257 &    -1.40 &     -.1507 &    -2.46 \\  
\\
   .62 &     1.5066 &    .0251 &    -1.40 &     -.2000 &    -2.47 \\  
   .64 &     1.4785 &    .0246 &    -1.41 &     -.2494 &    -2.47 \\  
   .66 &     1.4501 &    .0240 &    -1.42 &     -.2987 &    -2.47 \\  
   .68 &     1.4214 &    .0235 &    -1.44 &     -.3481 &    -2.47 \\  
   .70 &     1.3923 &    .0231 &    -1.45 &     -.3976 &    -2.47 \\  
\\
   .72 &     1.3632 &    .0226 &    -1.46 &     -.4470 &    -2.47 \\  
   .74 &     1.3342 &    .0222 &    -1.45 &     -.4964 &    -2.47 \\  
   .76 &     1.3054 &    .0217 &    -1.44 &     -.5459 &    -2.47 \\  
   .78 &     1.2771 &    .0214 &    -1.42 &     -.5953 &    -2.47 \\  
   .80 &     1.2493 &    .0210 &    -1.39 &     -.6449 &    -2.48 \\  
\\
   .82 &     1.2223 &    .0207 &    -1.35 &     -.6945 &    -2.48 \\  
   .84 &     1.1959 &    .0204 &    -1.32 &     -.7442 &    -2.49 \\  
   .86 &     1.1694 &    .0202 &    -1.33 &     -.7938 &    -2.48 \\  
   .88 &     1.1422 &    .0200 &    -1.36 &     -.8433 &    -2.47 \\  
   .90 &     1.1135 &    .0197 &    -1.43 &     -.8934 &    -2.51 \\  
\\
   .92 &     1.0829 &    .0195 &    -1.53 &     -.9444 &    -2.55 \\  
   .94 &     1.0496 &    .0192 &    -1.67 &     -.9935 &    -2.46 \\  
   .96 &     1.0130 &    .0189 &    -1.83 &    -1.0391 &    -2.28 \\  
   .98 &      .9744 &    .0186 &    -1.93 &    -1.0895 &    -2.52 \\  
  1.00 &      .9363 &    .0183 &    -1.91 &    -1.1544 &    -3.24 \\  
\\
\hline
\label{tab:spectr}
\end{tabular}
\end{center}
\end{table}

\end{document}